   \documentclass{aa}
   \usepackage{graphicx}
   
\begin{document}
   \title{
The 3-D ionization structure and evolution of NGC 7009 (Saturn Nebula)
 \thanks{Based on observations made with: ESO Telescopes at the La Silla 
Observatories (program ID 65.I-0524), 
and the NASA/ESA Hubble Space Telescope, 
obtained from the data archive at the Space Telescope Institute. 
Observing programs: GO 6117 (P.I. Bruce Balick), GO 6119 (P.I. Howard Bond) and GO 8390 (P.I. Arsen Hajian). 
STScI is operated by the association of Universities for Research in   
Astronomy, Inc. under the NASA contract  NAS 5-26555.  We extensively 
apply the photo--ionization code CLOUDY, developed at the Institute of
Astronomy of the Cambridge University (Ferland et al. 1998).}
 \author{ F. Sabbadin  \inst{1} \and M. Turatto \inst{1} \and E. Cappellaro \inst{2} \and S. Benetti \inst{1} \and R. Ragazzoni \inst{3,4}}
   \offprints{F. Sabbadin, sabbadin@pd.astro.it}
   \institute{INAF - Osservatorio Astronomico di Padova, vicolo dell'Osservatorio 5, I-35122 Padova, Italy \and 
   INAF - Osservatorio Astronomico di Capodimonte, via Moiariello 11, I-80131 Napoli, Italy \and INAF - Osservatorio 
   Astrofisico di Arcetri, Largo E. Fermi 5, I-50125, Italy \and Max-Planck-Institut f\"ur Astronomie, Koenigstuhl 17, D-69117 Heidelberg, 
   Germany}}
   \date{Received July 9, 2003; accepted September 11, 2006}
     
   \abstract{Tomographic and 3-D analyses for extended, emission-line objects are applied to long-slit ESO NTT + EMMI high-resolution 
spectra of the intriguing 
   planetary nebula NGC 7009, covered at twelve position angles. We derive the gas expansion law, the
   diagnostics and ionic radial profiles, the distance and the central
   star parameters, the nebular photo-ionization model and the spatial 
   recovery of the plasma structure and evolution. The Saturn Nebula (distance$\simeq$1.4 kpc, age$\simeq$6000 yr, 
ionized mass$\simeq$0.18 M$_\odot$) 
consists of several interconnected components, characterized by 
different morphology, physical conditions, excitation  and kinematics. We identify four ``large-scale'', medium-to-high excitation 
sub--systems (the internal shell, the main shell, the outer shell and the halo), and as many ``small-scale'' 
ones: the caps (strings of low-excitation knots within the outer shell), the ansae (polar, low-excitation, likely shocked 
layers), the streams (high-excitation polar regions connecting the main shell with the ansae), and an equatorial, medium-to-low excitation 
pseudo-ring within the outer shell. The internal shell, 
the main shell, the streams and the ansae expand at $V{\rm_{exp}}\simeq$4.0$\times$R\arcsec\, km s$^{-1}$, the outer shell, the caps and the 
equatorial 
pseudo-ring at 
$V{\rm_{exp}}\simeq$3.15$\times$R\arcsec\, km s$^{-1}$, and the halo at $V{\rm_{exp}}\simeq$10  km s$^{-1}$. 
We compare the radial distribution of the physical conditions and the line fluxes observed in the eight sub-systems with the theoretical 
profiles coming from the 
photo-ionization code CLOUDY, inferring that all the spectral characteristics of NGC 7009 are explainable in terms of photo-ionization by the  
central star, a hot (logT$_*$$\simeq$4.95) and luminous (log L$_*$/L$_\odot$$\simeq$3.70) 0.60--0.61 M$_\odot$ post--AGB star 
in the hydrogen-shell nuclear burning phase. The 3--D shaping of the Saturn Nebula is discussed within an evolutionary scenario dominated 
by photo-ionization and supported by the fast stellar wind: it begins with the superwind ejection (first isotropic, then polar deficient), 
passes through the neutral, transition 
phase (lasting $\simeq$ 3000 yr), the ionization start (occurred $\simeq$2000 yr ago), and the full ionization of the main shell ($\simeq$1000 
yr ago), at last reaching the present days: the whole nebula is optically thin to the UV stellar flux, except the caps (mean latitude  
condensations in the outer shell, shadowed by the main shell) and the ansae (supersonic ionization fronts along the major axis). 
   }
   
   \titlerunning{The Planetary Nebula NGC 7009}
   
   \maketitle
%
\section{Introduction} 

NGC 7009 (PNG 037.7-34.5, Acker et al. 1992) is a fairly bright, well-studied planetary nebula (PN) 
exhibiting large stratification effects.
It is also named the ``Saturn Nebula'' because of the peculiar optical appearance, as seen in Fig. 1, 
where the principal morphological sub-systems are indicated. The high-excitation He II emission forms a faint, elongated, irregular disc 
with an oval ring better defined out of the apparent major axis. In H$\alpha$ the deformed, hollow main shell (sometimes called ``the rim''; 
Balick et al. 1994, 1998) presents two faint streams 
spreading out along the major axis; it is surrounded by an outer shell and  
an even weaker, diffuse halo. The [N II] emission (bottom panel in Fig. 1; the gaps in the WFPC2 detectors are clearly 
visible) is concentrated in the external ansae along the major 
axis, and in the caps (i.e. two series of condensations not aligned with the major axis). 

In short: the main and outer shells, the halo 
and the streams connecting the main shell with the ansae are medium-to-high excitation regions, whereas the caps and the ansae 
present enhanced low-to-very low-excitation emissions. 
Spectacular HST multi-color reproductions of the Saturn nebula are 
in Balick et al. (1998) and Hajian \& Terzian at http://ad.usno.navy. mil/pne/gallery.html.

\begin{figure} \centering
\caption{WFPC2 (HST; programme GO 6117; P.I. Bruce Balick) appearance of NGC 7009 (logarithmic scale) 
in the high-excitation line $\lambda$4686  $\rm \AA\/$ of He II (top panel), in 
H$\alpha$ (which is emitted by the whole ionized nebula; central panel), and in the low-excitation line 
$\lambda$6584 $\rm \AA\/$ of [N II] (bottom panel; the image is spread in the four WFPC2 detectors). North is up and East to the left. The main 
morphological sub-systems are indicated.}  
\end{figure}

Based on high-dispersion spectroscopy, Weedman (1968) argues that the main shell of NGC 7009 consists of a thin prolate spheroid (an ellipse 
rotating around the major axis) seen perpendicular to the major axis. 

Fabry-Perot interferometry and narrow-band imaging by Reay \& Atherton (1985) indicate that the caps and the 
ansae constitute a double pair of condensations symmetrically displaced in velocity about the 
central star. The caps expand at $\simeq$38 km s$^{-1}$ (the western condensation is approaching and the 
eastern one receding); the line through the caps has an angle of 51$^o$ with the line of sight. The ansae move at $\simeq$60 
km s$^{-1}$ (the eastern one is approaching and the 
western one receding); the line of sight forms an angle of 84$^o$ with the line through the ansae, 
i.e. they almost lie in the plane of the sky.

Ground-based spectroscopy and HST multi--wavelength imagery by Balick et al. (1994, 1998) indicate that the caps and the ansae are 
genuine N-enriched  
FLIERs (fast, low-ionization emission regions; according to the definition by Balick et al. 1993), each cap consisting of a complex 
network of small knots. Although the 
``hammerhead'' appearance at the outer edge 
of the ansae is typical of a shock front, both pairs of FLIERs exhibit a decreasing ionization with 
distance from the star, as expected of a photo-ionized gas.

Also according to Hyung \& Aller (1995a, b) the observed spectrum of NGC 7009 is fairly well fitted by a 
photo-ionization model. They analyze two regions in the bright ring on the major and 
minor axes (dominated by the low-excitation and the high-excitation lines, respectively), obtain that the caps are N-rich knots 
and the ansae, 
very likely, do not lie along an axis almost perpendicular to the observer, and suggest a nebular model 
consisting of a dense toroid ring (or quasi-flat shell) + broad conical shells. 

In a very recent paper Gon\c calves et al. (2003) present low-resolution spectra along the major axis of NGC 7009, and 
discuss the physical excitation and chemical properties of the main shell, the caps, the streams and the ansae, obtaining that:

- all regions are mainly radiatively excited,

- there are no clear abundance changes across the nebula, but only marginal evidence for a modest overabundance of nitrogen in the ansae,

- none of the available theoretical models is able to account for the observed characteristics of the ansae.

Gon\c calves et al. (2003) also stress the need for an accurate kinematical study of the nebula. 

Deep [O III] imaging by Moreno-Corral et al. (1998) reveals the presence of a large (r$\simeq$130\arcsec), faint, complex  halo 
around NGC 7009, characterized by a series of fractured filaments, semi-circular envelopes and irregular condensations.

Our nebula is an extended X-ray emitting source (Guerrero et al. 2002): its central cavity consists of low-density (a few 
tens H-atom cm$^{-3}$), hot (T$\simeq$1.8$\times$10$^6$ K), shocked fast stellar wind. In fact, 
the exciting star is losing mass at high velocity, V$_{\rm edge}$$\simeq$2700 km s$^{-1}$ 
(Cerruti-Sola \& Perinotto 1985) and at an uncertain rate: 3.0$\times$10$^{-10}$ M$_\odot$ yr$^{-1}$ (Cerruti-Sola \& 
Perinotto 1985), 1.0$\times$10$^{-8}$ M$_\odot$ yr$^{-1}$ (Bombeck et al. 1986), 2.8$\times$10$^{-9}$ M$_\odot$ 
yr$^{-1}$ (Cerruti-Sola \& Perinotto 1989), 2.1$\times$10$^{-9}$ 
M$_\odot$ yr$^{-1}$ (Hutsemekers \& Surdej 1989), and 2.0$\times$10$^{-9}$ M$_\odot$ yr$^{-1}$ (Tinkler \& Lamers 
2002).

In spite of the great deal of attention, both theoretical and observational, and of the rich bibliography 
on NGC 7009, many (even fundamental) topics remain uncertain, e.g.: (a) the nebular distance (between 0.6 and 2.5 
kpc), and the central star mass (0.54 M$_\odot\le$M$_*\le$0.70 M$_\odot$) and luminosity (3.0$\le$log L$_*$/L$_\odot\le$4.2); 
(b) the excitation processes (shocks or photo-ionization) and the physical mechanisms 
(ionization-front instabilities, stellar or nebular bullets, photo-evaporation of neutral knots by a diluted, fast stellar wind, 
Oort--Spitzer rockets, i.e. photo-evaporation of neutral knots in the absence of a fast stellar wind; for details, see Balick et al. 1998) 
forming and shaping the caps and the ansae; 
(c) the discrepancy in the chemical abundances obtained from recombination and collisionally excited lines 
(Hyung \& Aller 1995b, Liu et al. 1995, Luo et al. 2001); 
(d) the overall, de-projected spatial structure of the ionized (and neutral) gas making up the Saturn Nebula. 

To deepen our insights in the kinematics, physical conditions, ionic and spatial structure, and evolutionary status 
of this fascinating object, NGC 7009 was included in the sample of 
proto-PNe and PNe observed with ESO NTT+EMMI (spectral range
$\lambda\lambda$3900-7900 $\rm\AA\/$, spectral resolution $\lambda$/$\Delta$$\lambda$=R=60\,000, spatial resolution S$\simeq$1.0\arcsec) and the 
Telescopio Nazionale Galileo (TNG)+SARG (spectral range  
$\lambda\lambda$4600-8000 $\rm\AA\/$, R=115\,000, S$\simeq$0.7\arcsec). The echellograms were 
reduced and analyzed according to 
the 3-D methodology so far applied to NGC 6565 (Turatto et al. 2002) and NGC 6818 (Benetti et al. 2003).

The results are contained in this paper, 
whose structure is as follows: Sect. 2 describes the 
observational procedure, the spectroscopic material and the reduction method, Sect. 3 disentangles the kinematics of the 
different sub--systems, Sect. 4 studies the radial profile of the physical conditions (electron 
temperature and electron density) in different directions, in Sect. 5 
we perform a critical analysis of the nebular distance, size, mass and age, in Sect. 6 the central star parameters are given,
Sect. 7 combines the observed ionic profiles and the photo-ionization model (CLOUDY),
Sect. 8 describes the 3-D spatio--kinematical structure of NGC 7009, and Sect. 9 contains the general discussion, sketches the 
shaping of the whole nebula and the different sub--systems, and draws the conclusions.

\section{Observations and reductions}
\begin{figure*}
   \centering 
   \caption{Detailed structure (on a logarithmic scale) of representative emission
at the twelve observed PA of NGC 7009. The original fluxes are multiplied by the factor given in parentheses, to render each emission 
comparable with $\lambda$5007 $\rm\AA\/$ of [O III]. The blue-shifted gas is to the left. 
The slit orientation at each PA is indicated in the He II frame. At PA=79$\degr$ (apparent major axis) the weak 
[S III] line at 
$\lambda$6312 $\rm\AA\/$ (echelle order 97) partially overlaps the $\lambda$6584 $\rm\AA\/$ [N II] emission (echelle order 93) of the W-SW 
ansa; see the text for details.}  
\end{figure*}
\begin{figure*}
   \centering 
   \caption{Same as Fig. 2.}  
\end{figure*}
To obtain an adequate spatial and spectral coverage of this extended and complex nebula, NGC 7009 was 
observed with ESO NTT + EMMI (echelle mode; grating $\#$14, grism $\#3$) at twelve equally spaced position 
angles (PA), including 
PA=79$\degr$, the apparent major axis containing the streams and the ansae. In all cases the spectrograph slit (1.0\arcsec \,wide 
and 60\arcsec\,long) was  
centered on the exciting star, whose continuum is used as position marker. During the observations 
the sky was photometric and the seeing fluctuated between 0.50\arcsec\,and 0.75\arcsec.

The echellograms (exposure times 360s to 600s) cover the spectral range  $\lambda$$\lambda$3900--7900 $\rm\AA\/$ with resolution 
R$\simeq$60\,000, and provide 
the spatio--kinematical structure of the main ionic species ([O I], ionization potential (IP)=0 eV
at $\lambda$6300 $\rm\AA\/$ to [Ar V], IP=59.8 eV at $\lambda$7005 $\rm\AA\/$) within the nebular slices covered by the 
spectrograph slit. 

Bias, zero--order flat field and distortion corrections, and wavelength and flux calibrations were performed according 
to the straightforward procedure described by Turatto et al. (2002). 
For educational purposes in Figs. 2 and 3 we present the spectral structure of NGC 7009 in each of the twelve observed PA at 
low, mean and high-excitation ($\lambda$6584 $\rm\AA\/$ of [N II], $\lambda$5007 $\rm\AA\/$ of [O III] and $\lambda$4686 $\rm\AA\/$ of He II, 
respectively), and at H$\alpha$ ($\lambda$6563 $\AA\/$ of H I), which is emitted through the whole of the ionized nebula, independent of the 
plasma excitation conditions. 

These pictures highlight the complex ionization structure of the Saturn Nebula: 
\begin{description} 
\item[-] the N$^+$ emission is faint everywhere, except in the caps and the ansae. The same is 
observed at even lower 
excitation, like S$^+$ at $\lambda$6717 $\rm\AA\/$ and $\lambda$6731 $\rm\AA\/$, and O$^o$ at $\lambda$6300 $\rm\AA\/$. 
Note that, due to the large optical extent of the nebula along the major axis (PA=79$\degr$), in Fig. 2 (bottom panel) the weak 
[S III] line at 
$\lambda$6312 $\rm\AA\/$ (echelle order 97) partially overlaps the $\lambda$6584 $\rm\AA\/$ [N II] emission (echelle order 93) of the W-SW 
ansa. Hereafter the net contribution of [N II] at this PA is obtained thanks to the similarity of 
the ionization potential of S$^{++}$ and Ar$^{++}$ (23.4 eV and 27.6 eV, respectively): we subtract the whole  
spectral image of [Ar III] at $\lambda$7135 $\rm\AA\/$ (properly scaled in flux) from that of [S III] at $\lambda$6312 $\rm\AA\/$, 
as in NGC 7009   
F($\lambda$7135 $\rm\AA\/$)/F($\lambda$6312 $\rm\AA\/$)$\simeq$10.0 (Kingsburgh \& Barlow 1994, Hyung \& Aller 1995a, b, this paper);
\item[-] H$\alpha$ is blurred by thermal motions, fine structure and expansion velocity gradient. 
Although this masks the detailed structure of the ionized gas, the H$^+$ distribution mimics that of O$^{++}$;
\item[-] $\lambda$5007 $\rm\AA\/$ of O$^{++}$, the strongest line in the 
optical region, represents the overall characteristics of the ionized gas (except the innermost regions, where 
O$^{+3}$ and 
higher-ionization species prevail, and the outer edge of the caps and the ansae, rich in O$^+$ and O$^o$);
\item[-] the He$^+$ $\lambda$4686 $\rm\AA\/$ line marks the highest-excitation nebular regions. The weak, blue-shifted, 
diffuse emission present at all PA is due to the tail in the thirteen fine-structure components of this recombination line (also 
broadened by thermal motions).
\end{description}

From Figs. 2 and 3 we infer that the main shell of NGC 7009 is a thin tri-axial ellipsoid, broken 
(or fractured) along 
the major axis (projected in PA=79$\degr$), where the streams and the ansae appear. Note the line tilt along the apparent 
minor axis 
(PA=169$\degr$), suggesting that we are not aligned with either the intermediate or the minor axes of the ellipsoid, and that 
the line of the nodes lies at PA$\simeq$130$\degr$--310$\degr$. 
The outer shell is an attached, complex, broad envelope completely filled with matter, as suggested by the absence of splitting at the 
rest wavelength of the [O III] line (out of the main shell). Kinematically the caps belong to the outer shell. 

Note that the main shell is red-shifted (by a few km s$^{-1}$) with respect to 
the outer shell (+ caps) and the halo; when considering the baricentric radial velocity of the ansae and the caps, we have: 
$V_{\rm rad}$(outer shell) $\simeq V_{\rm rad}$(ansae) $\simeq V_{\rm rad}$(caps) $\simeq V_{\rm rad}$(halo) 
$\simeq V_{\rm rad}$(main shell) - 3($\pm$1) km s$^{-1}$.
The streams at PA=79$\degr$ present hybrid kinematics: the internal region is co-moving 
with the main shell, the external one with the ansae (and the halo). 
Also note the variable FWHM along the streams, and the hooked appearance of the ansae (with both ``tips'' blue-shifted!).


Our preliminary, qualitative analysis stresses the complexity of NGC 7009. This appears to be a common characteristic of all PNe covered at 
adequate spatial and spectral resolutions, and questions the validity of the spatio-kinematical studies based on unsuitable observational 
material or, even worse, on the assumption of 
spherical symmetry for the emitting gas. Moreover, Figs. 2 and 3 can only provide 
a rough picture of the large amount of kinematical and physical information stored in the echellograms.  
Let's start our thorough investigation with the gas kinematics.
\section{The gas spatio-kinematics}

From Fig. 4,  
the spectrum of an elementary volume within the slice of nebula selected by the spectrograph slit is characterized by:
\begin{equation}
V_{\rm rad}= V_{\rm neb} - V_{\rm exp}\times {\rm cos\, a}  
\end{equation}
and
\begin{equation}
r= \vert R\times {\rm sin\, a} \vert, 
\end{equation}
where:
\begin{description}
\item[]$  V_{\rm neb}$ is the systemic radial velocity of the whole nebula; 
\item[]$ V_{\rm exp}$ is the expansion velocity; for PNe it is generally written in the form $V_{\rm exp}$=c$_1$$\times$R+c$_2$ (Wilson 1950, 
Weedman 1968);
\item[] R and r are the true and the apparent distances from the central star, respectively;
\item[] a is the angle between the line of sight and the central star-elementary volume direction.
\end{description}
Following Turatto et al. (2002, and references therein), for each spectral image of NGC 7009 we select: 
\begin{description}
\item[-] the ``central star pixel line'' (cspl), representing  the nebular gas projected at the apparent position of the star, whose 
motion is purely radial. The cspl is the same at all PA; in Eqs. (1) and (2) it corresponds to a=0$\degr$ and 180$\degr$; i.e. 
2$V_{\rm exp}$=$V_{\rm rad}$(a=180$\degr$) - $V_{\rm rad}$(a=0$\degr$), and r$_{\rm cspl}$=0; 
\item[-] the ``zero velocity pixel column'' (zvpc), which is unaffected by the expansion velocity field, and gives the spatial distribution 
of the ionized gas 
moving perpendicular to the line of sight (valid for a=90$\degr$ and 270$\degr$; i.e. $V_{\rm rad}$=$V_{\rm neb}$, and 
2r$_{\rm zvpc}$=2R$_{\rm zvpc}$ =R(a=90$\degr$) + R(a=270$\degr$)). 
\end{description}
We will derive the overall spatio-kinematical properties of the Saturn Nebula by assembling the kinematical results in a single radial direction 
(the cspl), and the spatial results in twelve tangential directions (the zvpc at each PA). 
Given the nebular complexity, the different sub-systems will be analyzed separately.
\begin{figure*} \centering
\caption{Nebular slice vs. spectral image connection for a typical PN. In this example we use 
the spectral image of the [O III] $\lambda$5007 $\rm\AA\/$ line of NGC 7009 (brightest layers of the main shell at PA=169$\degr$; see Fig. 3, 
bottom panel). The ``central star pixel line'' (cspl) and the ``zero velocity pixel column'' (zvpc) of the spectral image, as discussed in the 
text, are indicated.}  
\end{figure*}
\subsection{Kinematics of the main shell (+ streams + ansae)}
\begin{centering}
\begin{table*}
\caption{Peak separation in the cspl of NGC~7009 (main shell)}
\begin{tabular}{ccccccccccc}
\hline
\\
Ion &IP (eV)& $\lambda$ ($\rm\AA\/$)&&&&2$V_{\rm exp}$& (km/s)&\\
\cline {4-11}
&&& (1) & (2) & (3) & (4) & (5) & (6) & (7) & this paper\\
\\
\hline
\\
$[$S II$]$  & 10.4  & 6717-6731    &  -   &    -   &  -  & - & -  & -  & -   & 41.9  \\
$[$O II$]$  & 13.6  & 3727-7319       &40.9  & -      & 40.0& - & -  & -  & -   & 41.5:  \\
H I         & 13.6  & 6563       &42.1  & 44.0   &  -  &35.6 &32.4&36.8&34.5 &40.0  \\
$[$N II$]$  & 14.5  &  6584      &-     &   -    &  -  & - & -  & -  & -   &41.6  \\
$[$S III$]$ & 23.4  & 6312       &-     &      - &  -  & - & -  & -  & -   &41.4  \\
$[$Cl III$]$& 23.8  &5517-5537     &-     &      - &  -  & - & -  & -  & -   & 41.0: \\
C II        & 24.4  & 6578       &-     &      - &  -  & - & -  & -  & -   & 40.0: \\
He I        & 24.6  & 5876       &44.1  &      - &  -  & - & -  & -  & -   & 40.5 \\
$[$Ar III$]$& 27.6  & 7135       &-     &  -     &    -& - & -  & -  & -   & 40.8   \\
$[$O III$]$ & 35.1  & 5007       &41.1  &  -     &41.2 & - & -  & -  & -   &40.8  \\
$[$Ar IV$]$ & 40.7  & 4711-4740    &39.7  &    -   & -   & - & -  & -  & -   & 39.7 \\
$[$Ne III$]$& 41.0  & 3869-3967       &38.7  &      - &  -  & - & -  & -  & -   & 40.0 \\
N III       & 47.4  & 4640       & 41.1 &      - &  -  & - & -  & -  & -   & 40.0: \\
He II       & 54.4  & 4686       &35.8  &      - &34.6 & - & -  & -  & -   & 36.6 \\
O III       & 54.9  & 3444-3770       &32.9  &      - &  -  & - & -  & -  & 30.5& - \\
$[$Ar V$]$  & 59.8  & 7005       &-     &      - &  -  & - & -  & -  & -   & 33.0: \\
\\
\end{tabular}
\\
\begin{tabular}{lllll}
\\
(1)=& Wilson (1950)&  (5)=& Cristiani et al. (1989)\\
(2)=& Mendez et al. (1988)& (6)=& Bianchi (1992)\\
(3)=& Meatheringham et al. (1988)& (7)=& Liu \& Danziger (1993)\\
(4)=& Bianchi et al. (1989)\\
\end{tabular}
\\
\\
\end{table*}
\end{centering}

The peak separations in the cspl, 2$V_{\rm exp}$, are contained in Table 1 (last column), where the
ions are put in order of increasing IP ([O I] at $\lambda$6300 $\rm\AA\/$ is absent, since the main shell is a medium-to-high excitation region). 
Typical errors are 1.0 km s$^{-1}$ for the strongest forbidden emissions (like
$\lambda$5007 $\rm\AA\/$ of [O III])
to 2.0 km s$^{-1}$ for the faintest ones (in particular, 
$\lambda$5517 $\rm\AA\/$  and $\lambda$5537 $\rm\AA\/$ of [Cl III], $\lambda$6578 $\rm\AA\/$ of C II and $\lambda$7005 $\rm\AA\/$ of [Ar V]). 
The errors for the recombination lines
are: 2.0 km s$^{-1}$ for $\lambda$6563 $\rm\AA\/$ of H I, and 1.5 km
s$^{-1}$ for $\lambda$5876 $\rm\AA\/$ of He I and $\lambda$4686
$\rm\AA\/$ of He II. 
Table 1 also contains the kinematical results taken from  
the literature. 

Concerning the zvpc in the main shell, the intensity peak separations, 2r$_{\rm zvpc}$, 
in the different emissions at the twelve PA of NGC 7009 are listed in Table 2. 
Please note the different ``relative'' spectral and spatial resolutions of our echellograms: the former is   
given by $V_{\rm exp}$/$\Delta$V$\simeq$4 ($\Delta$V=spectral resolution), and the latter by 
r/$\Delta$r $\simeq$8--16 (r=apparent radius, $\Delta$r=seeing); this means that the 
spatial information of NGC 7009 is more accurate than the kinematical information.

\begin{centering}
\begin{table*}
\caption{Peak separation in the zvpc at the twelve PA of NGC~7009 (main shell)}
\begin{tabular}{ccccccccccccc}
\hline
\\
Ion&&&&&2r$_{\rm zvpc}$&(arcsec)&&&&&&\\ 
\cline{2-13}
&PA=04$\degr$ &19$\degr$ &34$\degr$ &49$\degr$ &64$\degr$ &79$\degr$ 
&94$\degr$ &109$\degr$&124$\degr$&139$\degr$&154$\degr$&169$\degr$  \\
\\
\hline 
\\  
$[$S II$]$    & 12.1   & 13.4  & 14.4  & 18.5  & 23.5 & 24.8 &19.4 & 13.6 & 11.4&10.6  &10.6  &10.5  \\
$[$O II$]$    & 12.0   & 13.2  & 14.2  & 18.0: & -    & -    &-    & 13.4 & 11.2&10.4  &10.4  &10.3 \\
HI            & 11.2   & 12.8  & 13.6  & 17.0  & 22.5 & 24.5 &18.5 & 12.7 & 10.7&10.0  &10.0  &10.2  \\
$[$N II$]$    & 12.1   & 13.3  & 14.3  & 18.3  & 23.5 & 24.7 &19.3 & 13.4 & 11.3&10.5  &10.4  &10.4  \\
$[$S III$]$   & 11.9   & 13.1  & 14.3  & 18.1  & 23.3 & 24.5 &19.0 & 13.3 & 11.2&10.3  &10.3  & 10.4 \\
$[$Cl III$]$  & 12.0:  & 13.0: & 13.8: & 18.2: & -    &-     & -   & 13.0:& -   &-     &10.3: &10.4: \\
C II          & 11.5:  & 13.1  & 14.1: & 18.0: & -    &-     & -   & 12.5:&10.8:&10.0: & -    &10.4:  \\
He I          & 12.0   & 13.0  & 14.1  & 18.0  & 23.5 & 24.5 & 18.7& 13.2 &11.2 &10.3  &10.3  &10.4  \\
$[$Ar III$]$  & 12.0   & 13.2  & 14.0  & 18.1  & 23.4 & 24.5 & 18.5& 13.2 &11.2 &10.3  &10.2  &10.3  \\
$[$O III$]$   & 12.0   & 13.2  & 14.1  & 18.0  & 23.3 & 24.5 & 18.3& 13.0 &11.2 &10.3  &10.3  &10.2  \\
$[$Ar IV$]$   & 11.3   & 12.8  & 13.8  & 17.7  & -    & -    & -   & 12.1 &10.8 &10.1  &9.9   &9.8   \\
$[$Ne III$]$  & 11.3   & 12.8  & 13.8  & 17.7  & 23.0 & 24.4 & 18.0& 12.0 &10.7 &10.1  &10.0  &9.9  \\
N III         & 11.1   & 12.6  & 13.3: & 17.3  & -    & -    & -   & 11.6:&10.0:& 9.9: &9.9   &9.8:   \\
He II         & 10.3   & 12.1  & 12.7  & 14.5  & 16.0:& -    & -   & 9.6  &9.3  & 9.2  &8.9   &9.1  \\ 
$[$Ar V$]$    & 8.8:   & -     & -     & -     & -    & -    & -   &  -   &  -  & 8.5: &8.3:  &8.3:  \\
\\
\hline
\end{tabular}
\end{table*}
\end{centering}

According to Benetti et al. (2003, and references therein), the next step is the identification of the PA at which 
R$_{\rm zvpc}$$\simeq$R$_{\rm cspl}$. 
Let us assume for the main shell of NGC 7009 the most 
general spatial structure, i. e. a tri-axial ellipsoid. The major axis, projected in PA=79$\degr$, is almost 
perpendicular to the line of sight, since the emissions appear nearly un-tilted. The intermediate and the minor axes lie close to 
PA$\simeq$169$\degr$; actually, the line-tilt observed here suggests that we are not aligned with either axis (see Fig. 4). We conclude that  
R(minor axis)$<$R$_{\rm zvpc}$$\simeq$R$_{\rm cspl}$$<$R(intermediate axis) along the apparent minor axis of the nebula, i.e. at  PA=169$\degr$.
 
The cspl--zvpc trend for the main shell of NGC 7009 at PA=169$\degr$, shown in Fig. 5, follows the Wilson law: the high-excitation 
particles expand more slowly than the low-excitation ones. Moreover, there is a positive correlation between the expansion velocity and 
the size of the monochromatic image (Wilson 1950). 
Note that, in spite of the wide range of IP observed 
(10.4 eV of S$^+$ to 59.8 eV of Ar$^{+4}$), the range of both 2r$_{\rm zvpc}$ and 2$V_{\rm exp}$ is quite limited (8.3 to 10.5 arcsec 
and 33.0 to 41.9 km s$^{-1}$, respectively). When added to the distribution of the data in Fig. 5, this suggests that NGC 7009 
at PA=169$\degr$ is optically thick in Ar$^{+4}$ and He$^{++}$, nearly thick in N$^{+3}$, Ne$^{+3}$ and Ar$^{+3}$, and thin in the other 
ionic species, 
and/or that the radial density profile of the main nebula is quite sharp. To gain insight into the question, a  kinematical study at 
even higher ionization states appears indispensable, e.g. in the forbidden line of Ne$^{+4}$ 
at $\lambda$3425 $\rm\AA\/$ (which, unfortunately, is outside our spectral range).

\begin{figure} \centering
\caption{The complex kinematics of NGC 7009 at PA= 169$\degr$ (apparent minor axis). 
Empty circles= zvpc vs cspl in the main shell, providing $V_{\rm exp}({\rm main\ shell})$ (km s$^{-1}$)= 
4.0$(\pm0.3)$$\times$R\arcsec; dots= extent of both the zvpc and the cspl at 
10$\%$ maximum intensity; this is valid for the outer shell and gives $V_{\rm exp}({\rm outer\ shell})$ (km s$^{-1}$)= 
3.15$(\pm0.3)$$\times$R\arcsec; moreover, $V_{\rm exp}$(halo)$\simeq$10 km s$^{-1}$. Long-dashed line= expansion law from Weedman (1968); 
continuous line= expansion law adopted in this paper.}  
\end{figure}


The r$_{\rm zvpc}$ vs IP trend being the same at all observed PA
(from Table 2), we can assess that the expansion velocity in the main shell of NGC 7009 is proportional to the distance 
from the central star through the relation:

\begin{equation}
  V_{\rm exp}({\rm main\ shell}) ({\rm km\ s^{-1}}) = 4.0 (\pm0.3)\times R\arcsec. 
\end{equation}

Weedman (1968) has obtained a different law: $V_{\rm exp}$(km s$^{-1}$)= $7.1\times(R\arcsec - 6.9)$ (also shown 
in Fig. 5), which appears questionable, since it is based on: (I) spectra at a single PA (along the apparent major axis); (II) the 
``a priori'' assumption that 
the nebula is a prolate spheroid with a=27.7\arcsec\, and a/b=1.5, seen perpendicular to the major axis.

Although weakened by projection effects, Eq. (3) is extended to the streams and 
the ansae, for reasons of continuity (Figs. 2 and 3). Both these sub--systems 
appear aligned to the major axis of the central ellipsoid, and highly inclined with respect to the line of sight 
(the resulting supersonic velocity of the ansae, $V_{\rm exp}$(ansae)$\simeq$100 km s$^{-1}$, is quite uncertain). 
Note the large, variable FWHM along the streams, and the hammerhead appearance of the ansae, suggestive of local turbulent motions (i.e. interaction 
with the slower outer shell and halo). 
 
\subsection{Kinematics of the outer shell (+ caps)}
The foregoing analysis cannot be applied to the outer shell, the cspl of NGC 7009 being dominated by the 
bright main shell. 
In order to derive the spatio-kinematical properties of the outer shell, we measure the extent of both the cspl and the zvpc 
at 10$\%$ maximum intensity (this cut avoids the contribution of the halo to the zvpc) in the strongest lines ($\lambda$5007 
$\rm\AA\/$ of [O III], $\lambda$5876 $\rm\AA\/$ of He I, $\lambda$6584 $\rm\AA\/$ of [N II] and $\lambda$7135 $\rm\AA\/$ of [Ar III]) at PA=169$\degr$ 
(the apparent minor axis). 
The results are graphically shown in Fig. 5. When extended to the other PA, and combined with the whole emission structure of 
the outer shell, they furnish the following expansion law:

\begin{equation}
  V_{\rm exp}({\rm outer\ shell})({\rm km\ s^{-1}}) = 3.15 (\pm0.3)\times R\arcsec, 
\end{equation}
valid for the caps too.

\subsection{Kinematics of the halo}
Only at PA=64$\degr$ and 94$\degr$ does the faint [O III] $\lambda$5007 $\rm\AA\/$ emission of the halo 
present a two-peaks velocity structure, providing $V_{\rm exp}$(halo) 
= 12($\pm$2) km s$^{-1}$, whereas it is not split at the other PA, with FWHM$\simeq$$2V{\rm exp}$(halo) 
= 20($\pm$5) km s$^{-1}$. The line split observed at PA=64$\degr$ and 94$\degr$ (i.e. close to the apparent major axis) is connected to the 
extended emissions surrounding 
the streams and the ansae (see  Fig. 1), which are reminiscent of bow shocks (Balick et al. 1998).

\subsection{Kinematics overview}

The intriguing kinematical properties of the Saturn Nebula are presented in Fig. 6, 
showing the position--velocity
(P--V) maps, i.e. the complete radial velocity field at the twelve
observed PA. We have selected He II, [O III] and [N II] as markers of the high, medium and low-excitation
regions, respectively. 
These  P--V maps are relative to the systemic heliocentric velocity
of the main shell, $V_{\rm rad \odot}$(main shell)= -46.0($\pm1.0$) km s$^{-1}$, corresponding
to $V_{\rm LSR}$(main shell)= -36.3($\pm$1.0) km s$^{-1}$, and are scaled according to
Eq. (3), i.e. they reproduce the tomographic maps of the main shell (+ the streams + the ansae) in the
nebular slices covered by the slit.  

Concerning the outer shell (+ the caps), the tomographic maps in Fig. 6 are:
\begin{description}
\item[(a)] compressed along the horizontal axis, since the kinematics of these sub--systems is provided by Eq. (4); this is 
particularly evident along and close to the apparent minor axis;
\item[(b)] slightly blue-shifted, as $V_{\rm rad \odot}$(outer shell)$\simeq$ $V_{\rm rad \odot}$(caps) $\simeq$ -49 km s$^{-1}$. 
\end{description}
Note the peculiar behaviour of the ansae: they follow the expansion law of the main shell and the streams, whereas their baricentric 
radial velocity is the same as that of the outer shell, the caps and the halo.

Fig. 6 highlights the main limitation of any spatio-kinematical study of the Saturn Nebula: the major axis (containing the most 
interesting features, i.e. the streams and the ansae) almost lies and moves in the plane of the sky, the radial component of the 
expansion velocity is nearly null, and projection effects play the leading role. In other words: we are forced to extrapolate the 
(well-defined) kinematical and spatial properties of the matter perpendicular to the major axis to the poor kinematics and 
precise spatial profile of the gas along the major axis. 

The P--V maps (Fig. 6) stress the variety of kinematics in the three ``large-scale'' sub--systems of 
the Saturn Nebula (the main shell, the outer shell and the halo) 
and in the three ``small-scale'''' ones (the streams, the caps and the ansae). Different evolutionary scenarios can qualitatively 
explain the ``large-scale'' features: (a) a multiple (triple) ejection, the halo being the oldest and the main shell the youngest; 
(b) different accelerating mechanisms acting on a single, prolongated ejection; 
(c) a combination of (a) and (b). 
To disentangle these possibilities (and their correlation with the ``small-scale'' structures) 
is premature, requiring more input, and is post--poned to Sects. 5 to 9. We wish to remark here that the common expansion law of the 
outer shell and the caps suggests that they are coeval; very likely, the latter simply represent local condensations within the 
former (a similar result is obtained by Corradi et al. 2000 for the enhanced low-excitation layers of IC 2553 and NGC 5882). At present we 
cannot assess that the same conclusion is valid for the 
three other co-moving sub--systems of NGC 7009 (i.e. main shell, 
the stream and the ansae). 

Besides the complexity of the kinematics, Fig. 6 highlights the large stratification of the radiation, the Saturn Nebula being 
optically thin in most directions. The ionization level decreases (i.e. the [N II] emission enhances) in:
\begin{description}
\item[-] two series of knots in the outer shell at PA=64 to 94$\degr$ (i.e. the caps); this appears connected to the shadowing by an inner, 
dense layer (the main shell);
\item[-] the ansae in PA=79$\degr$; probably associated to the presence of internal, extended, low-density regions along the major axis 
(the streams).
\end{description}
\begin{figure*} \centering
\caption{Position--velocity maps at the twelve observed PA of NGC 7009 for the high (He II, blue), medium ([O III], 
green) and low ([N II], red) excitation regions, scaled according to the relation $V_{\rm exp}$(km s$^{-1}$)=4.0$\times$R\arcsec \,(valid for 
the main shell + the streams + the ansae; see the text for details). The orientation of these tomographic maps is as in Figs. 2 and 3.}  
\end{figure*}
A final remark concerns the overall structure of the zvpc at all PA, providing the matter distribution in the plane of the sky passing through 
the central star. We 
discard H$\alpha$ as reference emission, because of the large broadening, 
and select $\lambda$5007 $\rm\AA\/$ of [O III], i.e. the strongest line 
of an ionic species present throughout the whole nebula (in first approximation we can put: 
I($\lambda$5007 $\rm\AA\/$)$_{\rm zvpc}\propto N_{\rm e}\, ^2$, 
$N_{\rm e}$ being the electron density). As shown in Fig. 7, all the sub--systems of NGC 7009 are present (except the caps, 
whose contribution in the zvpc is null, due to projection effects; see Figs. 2 and 3). 
Moreover, a further, internal, faint and incomplete shell appears 
close to the major axis (PA=49$\degr$ to PA=109$\degr$). As expected, this feature (hereafter called ``the internal shell'') is also 
present in the zvpc of [Ar III] and, even more strongly, He II, but it is absent (or almost absent) in those of [N II] and [S II], i.e. it is a 
high-excitation region. Concerning the expansion law of the internal shell, we can only say that it mimics that of the main shell.
\begin{figure}  \centering
\caption{Overall structure of the [O III] zvpc at the twelve observed PA of NGC 7009. The slit orientation is indicated; moreover: 
short-dashed line=internal shell; continuous line=main shell; long--dashed line=outer shell.}
\end{figure}

In summary, our detailed spatio-kinematical analysis confirms and extends the morphological complexity sketched in Sect. 1, NGC 7009 consisting 
of a puzzle of interconnected sub-systems, as synthesized in Table 3.

\begin{centering}
\begin{table*}
\caption{NGC 7009: summary of the kinematical and physical conditions in the sub--systems.}
\begin{tabular}{ccccccc}
\hline
\\
Sub-system&$V_{\rm exp}$ &$V{\rm rad}_{\odot}$ &$T_{\rm e}$   & $N_{\rm e}$ peak & excitation & first\\
          & (km s$^{-1}$)&         (km s$^{-1}$)&                              (K)  & (cm$^{-3}$)   &  degree & citation\\
\\
\hline
\\
internal shell&4.0$\times$R\arcsec (:)    &  -46 (:) &$\ge$12000  &  2000-2500 (:)  & very high &Sect. 3.4     \\
main shell   &  4.0$\times$R\arcsec   &  -46 & 12000 to 10000  &4000 to 8000& high&Sect. 1      \\
streams      &  4.0$\times$R\arcsec   &  -46 (inner) to -49 (outer)  & 10000 (:)    &  1000--1500& high& Sect. 1\\
ansae        &  4.0$\times$R\arcsec   &  -49 & 10000 (:)   &  2000--2500& low to very low &Sect. 1     \\
outer shell  &  3.15$\times$R\arcsec  &  -49 & 9500-10000   &  2000-2500& high &Sect. 1      \\
caps         &  3.15$\times$R\arcsec  &  -49  & 9500-10000   &  3000-3500& low to very low&Sect. 1  \\
equatorial pseudo-ring& 3.15$\times$R\arcsec  &-49  &10000&2800&mean to low&Sect. 8\\
halo         &  10                 &  -49  &  10000 (:) & 250-300  & high &Sect. 1   \\
\end{tabular}
\\
\\
\end{table*}
\end{centering}

\section{The physical conditions}
First of all the observed line intensities must be corrected for interstellar absorption according to:
\begin{equation}
\log \frac{{\rm I}(\lambda)_{\rm corr}}{{\rm I}(\lambda)_{\rm obs}}=f_{\lambda}\, c({\rm H}\beta),
\end{equation}
where f$_{\lambda}$ is the interstellar extinction coefficient given by Seaton (1979).  

The H$\alpha$/H$\beta$ analysis of
the whole spectral image, as introduced by Turatto et al. (2002), provides poor spatial resolution maps, because of the blurred 
appearance of both  H$\alpha$ and  H$\beta$. The resulting maps appear quite uniform at  H$\alpha$/H$\beta$=3.20  ($\pm$0.10), 
corresponding to  c(H$\beta$)=0.15 ($\pm$0.05) (for the case B of  Baker \& Menzel 1938, $T_{\rm e}$=10\,000 K and 
log~$N_{\rm e}$= 3.60; Brocklehurst 1971, Aller 1984, Hummer \& Storey 1987), with a smooth decline in the innermost regions (very likely caused 
by  a local increase of $T_{\rm e}$).

This agrees with most of the c(H$\beta$) estimates reported in the literature (0.09--0.27, Perinotto \& Benvenuti 1981; 0.39--0.56, 
Tylenda et al. 1992; 
0.09, Kingsburgh \& Barlow 1994; 0.05--0.28, Bohigas et al. 1994; 0.20, Liu et al. 1995; 0.09, Hyung \& Aller 1995a,b; 0.24, Lame \& Pogge 
1996; 0.14, Ciardullo et al. 1999; 0.07, Luo et al. 2001; 0.10, Rubin et al. 2002; 0.16, Gon\c calves et al 2003), indicating that NGC 7009 
is a PN with little absorption. 

The radial profile of the electron temperature, $T_{\rm e}$, is obtained from diagnostic line ratios 
($\lambda$5007 $\rm\AA\/$/$\lambda$4363 $\rm\AA\/$ of [O III] and $\lambda$6584 $\rm\AA\/$/$\lambda$5755 $\rm\AA\/$ of [N II]), and the 
$N_{\rm e}$ radial distribution from both line ratios ($\lambda$6717 $\rm\AA\/$/$\lambda$6731 $\rm\AA\/$ of [S II] and 
$\lambda$4711 $\rm\AA\/$/$\lambda$4740 $\rm\AA\/$ of [Ar IV]) and the absolute H$\alpha$ flux. 
We consider the zvpc, corresponding to the gas in the plane of the sky, since it is generally independent on the expansion velocity field 
(for details, see Turatto et al. 2002). 
In the specific case of NGC 7009 both methods (i.e. diagnostic line ratios and H$\alpha$ flux) suffer some limitations due to:
\begin{description}
\item[-] the weakness of the [N II] auroral line and the [S II] and [Ar IV] doublets. $T_{\rm e}$[N II], $N_{\rm e}$[S II] 
and $N_{\rm e}$[Ar IV] can be derived only at the intensity peaks. To get out of this trouble we extend the analysis 
of the diagnostic line ratios to the prominent knots of the whole spectral images;
\item[-] the large H$\alpha$ broadening. The deconvolution for instrumental resolution plus thermal motions plus fine structure is 
quite complex, and the resulting F(H$\alpha)_{\rm zvpc}$ and $N_{\rm e}$(H$\alpha$) profiles uncertain. Moreover, the different 
kinematical properties of the sub--systems must be considered, being :
\end{description}
\begin{equation}
N_{\rm e}(H\alpha)\propto \frac{1}{T_{\rm e}^{-0.47}} \times 
(\frac{F(H\alpha)_{\rm zvpc}}{\epsilon_{\rm l} \times r_{\rm cspl} \times D})^{1/2},
\end{equation}
where:
\begin{description}
\item[-] D is the nebular distance;
\item[-] r$_{\rm cspl}$ is the radius of the cspl in the main shell (this is valid for the internal shell, the main shell, the streams and the 
ansae) and in the outer shell (for the outer shell and the caps);
\item[-] $\epsilon_{\rm l}$ is the ``local filling factor'', representing the fraction of the local 
volume actually filled by matter with density $N_{\rm e}$. 
\end{description} 
Thus, according to Benetti et al. (2003), F(H$\alpha)_{\rm zvpc}$ 
is also obtained from the observed radial ionization structure relative to O$^{++}$  and the fair assumption that O/H=constant 
across the nebula (Sect. 7); in the $N_{\rm e}$ range here considered:
\begin{equation}
\frac{F(H\alpha)_{\rm zvpc}}{F(\lambda 5007\rm\AA)_{\rm zvpc}}\propto\frac{H}{O}\times f(T_{\rm e})\times icf(O^{++}),
\end{equation}

with $icf(O^{++})$=$\frac{O}{O^{++}}$=ionization correcting factor.

A comparative analysis gives satisfactory results: 
the $N_{\rm e}$(H$\alpha$) profiles derived in the two ways 
differ by less than 7$\%$, with the exception of the faint, innermost regions, where discrepancies as large as 30$\%$ are observed. 
At present we are unable to identify the principal factor responsible (inadequate deconvolution?  $T_{\rm e}$ uncertainties? chemical composition 
gradient?), although the last possibility is strongly suspected, due to the different nebular morphology in He$^{++}$ and H$^+$ (Fig. 1); this is 
discussed further in Sect. 7.

The resulting $N_{\rm e}$ radial profile rapidly changes with PA, as expected of a chaotic object like NGC 7009. Some 
representative examples are shown in Fig. 8. 
In order to cover all the sub--systems, we take into account:
\begin{description}
\item[-] the zvpc at PA=169$\degr$ for the main shell, the outer shell and the halo along the apparent minor axis,
\item[-] the zvpc at PA=79$\degr$ for the internal shell, the main shell, the streams and the ansae along the apparent major axis. 
In the following this direction 
will be considered as the true major axis of the nebula, due to the modest line tilt observed at PA=79$\degr$ (Figs. 2 and 6).
\end{description}
Concerning the caps, whose contribution to the zvpc is null because of perspective effects (Sects. 2 and 3), we first de--project the spectral 
images at PA=94$\degr$ (through Eq. (3) for the main shell, and Eq. (4) for the outer shell and the caps), and then 
consider the  direction containing the central star and the caps  (it forms an angle of 55$\degr$ with the line of sight).

Let us analyze the physical conditions in the different sub-systems (a synthesis is given in Table 3). 

\subsection{The internal shell}
We could derive only poor information for this elusive, high-excitation sub-system: it is a moderate-density, $N_{\rm e}$(peak)$\simeq$2000-2500 
cm$^{-3}$, thin region at large $T_{\rm e}$ ($\ge$12\,000 K).

\subsection{The main shell}

$T_{\rm e}$[O III] presents a well-defined radial profile common at all PA: it is $\ge$12\,000 K in the 
tenuous, innermost regions, it rapidly decreases outward down to 10\,000 K in the densest layers, and 
furtherout it remains more or less constant, in quantitative agreement with the results by Rubin et al. (2002), based on HST/WFPC2 imagery.  
Ground--based $T_{\rm e}$[O III] determinations (mean value in the brightest regions) by Kingsburgh \& Barlow (1994), Bohigas et 
al. (1994), Hyung \& Aller (1995a, b), Liu et al. (1995), Mathis et al. (1998), Kwitter \& Henry (1998), Luo et al. (2001) and Gon\c calves et al. 
(2003) cluster around 10\,000 K. 

The $N_{\rm e}$ radial distribution in the main shell shows a sharp, bell-shape profile with $N_{\rm e}$[S II] peaks up to:
\begin{description}
\item[] 4000  ($\pm$500) cm$^{-3}$ along the major axis, 
\item[] 7000 ($\pm$500) cm$^{-3}$ along and close to the minor axis, 
\item[] 8000 ($\pm$500) cm$^{-3}$ at PA=94$\degr$, in the de-projected direction containing the caps.
\end{description}
Previously, Hyung \& Aller (1995a, b) obtained $N_{\rm e}$[S II] peaks of 5000 cm$^{-3}$ and 5600  cm$^{-3}$ along the apparent major 
and minor axes, respectively, and Gon\c calves et al. (2003)  $N_{\rm e}$[S II]=5500--5900 cm$^{-3}$ 
along the apparent major axis (cf. Sect. 4.6).

$N_{\rm e}$[Ar IV] is systematically lower than $N_{\rm e}$[S II]. We adopt the [Ar IV] electron impact excitation rates 
given by Keenan et al. (1997); when using earlier collisional rates (for example, Aller 1984), $N_{\rm e}$[Ar IV]$\simeq N_{\rm e}$[S II].

In general we have: $N_{\rm e}$[S II] $\times \epsilon_{\rm l}^{0.5}\simeq N_{\rm e}$(H$\alpha)$ (Aller 1984, Pottasch 1984, 
Osterbrock 1989). 
In Fig. 8 the match between $N_{\rm e}$(forbidden lines) and $N_{\rm e}$(H$\alpha$) occurs for 
(r$_{\rm cspl}\times$D$\times\epsilon_{\rm l})_{\rm main\ shell}$$\simeq$5.0 arcsec kpc, 
where r$_{\rm cspl}$=size of the main shell in the radial direction, and D=nebular distance (Eq. (6); for details, see Turatto et al. 2002), i.e. 
(D$\times\epsilon_{\rm l})_{\rm main\ shell}$ $\simeq$1.0 kpc (from Table 2).

\subsection{The streams}
No direct $T_{\rm e}$ determination can be obtained for these medium-to-high excitation, low-density ($N_{\rm e}$$\simeq$1000--1500 cm$^{-3}$) regions 
connecting the main shell with the ansae, due to the weakness of the [O III] auroral line. Following Rubin et al. (2002) 
and Gon\c calves et al. (2003), we adopt $T_{\rm e}\simeq$ 10\,000 K.

\subsection{The ansae}
We derive only a rough estimate of $T_{\rm e}$[N II]$\simeq$ 10\,000 K. Balick et al. (1994) report 
$T_{\rm e}$[O III]=11\,500 K and $T_{\rm e}$[N II]=8100 K, Kwitter \& Henry (1998) $T_{\rm e}$[O III]=9300 K and $T_{\rm e}$[N II]=9100 K, 
Rubin et al. (2002) $T_{\rm e}$[O III]$\simeq$9200 K, and Gon\c calves et al. 
(2003) $T_{\rm e}$[O III]=9300-10\,100 K. Likely, $T_{\rm e}$ decreases at 
the outer edge, where the ionization drops.
$N_{\rm e}$[S II] reaches 2000--2500 cm$^{-3}$ in the ansae, i.e. they hardly double the density of the streams, in agreement with 
Balick et al. (1994, 1998), and in partial agreement with Gon\c calves et al. (2003), who find $N_{\rm e}$[S II](streams) $\simeq 
N_{\rm e}$[S II](ansae)=1300--2000 cm$^{-3}$.

\subsection{The outer shell} Our echellograms provide $T_{\rm e}$[O III]$\simeq$9500-10\,000 K and  $N_{\rm e}$[S II] up to 2000-2500 cm$^{-3}$. 
Previously, Kwitter \& Henry (1998) obtained $T_{\rm e}$[O III]=9400 K and $T_{\rm e}$[N II]=12\,100 K, and Rubin et al. (2002) 
$T_{\rm e}$[O III]$\simeq$9200 K. Note the presence of a localized knot with $N_{\rm e}$[S II]$\simeq$ 2800 cm$^{-3}$ in the northern sector of 
PA=169$\degr$ (see the [N II] panel in Fig. 1). 
The match of $N_{\rm e}$(forbidden lines) and $N_{\rm e}$(H$\alpha$) provides (r$_{\rm cspl}\times$D$\times\epsilon_{\rm l})_{\rm outer\ shell}$ 
$\simeq$10--12 arcsec kpc 
(r$_{\rm cspl}$= size of the outer shell in the radial direction), implying that  
(D$\times\epsilon_{\rm l})_{\rm outer\ shell}$ $\simeq$1.0 kpc, i.e. the same value derived for the main shell (Sect. 4.2).

\subsection{The caps} 
In these medium-latitude condensations of the outer shell $T_{\rm e}$[O III]$\simeq T_{\rm e}$[N II]$\simeq$ 10\,000 K,  and 
$N_{\rm e}$[S II] $\simeq$3000-3500 cm$^{-3}$; similar 
results are obtained by Balick et al. (1994), Kwitter \& Henry (1998) and Rubin et al. (2002), whereas, according to Gon\c calves et al. 
(2003), $T_{\rm e}$[O III]$\simeq T_{\rm e}$[N II]$\simeq$9300--10\,400 K, and $N_{\rm e}$[S II]=4500--5000 cm$^{-3}$. The large density reported 
by Gon\c calves et al. appears incompatible with the weakness of the caps in the H$\alpha$ 
panel of Fig. 1 (since, in first approximation, I(H$\alpha)$$\propto$$N_e\, ^2$). The same occurs when considering both the H$\alpha$ 
and [O III] spectral 
images at PA=79$\degr$ (Fig. 2, bottom panel). Our conclusion is that, due to projection effects, the line intensities provided 
by the low-resolution spectra of 
Gon\c calves et al. (2003; positions K2 and K3 in their Table 1) represent a mix of both the caps and the underlying main shell, which is denser 
and at higher excitation (as 
confirmed by the analysis of our echellograms at PA=79$\degr$ in ``low spectral resolution'' mode, i.e. integrated along the expansion 
velocity field). 
This is a clear 
demonstration of the severe limits implicit in the study of objects as complex as NGC 7009 at unsuitable spectral resolution. On the contrary, the 
tomographic and 3-D methodologies, based on high-resolution spectra, reconstruct the true spatial structure of the gas and overcome any misleading 
camouflage due to projection. 
We stress that the matter distribution in Fig. 8 (bottom 
panel) enforces the indication outlined by the kinematics (Sect. 3.4): the low-excitation degree of the caps is likely due to the shadowing by 
the main shell, a triaxial ellipsoid 
of large ellipticity, quite dense along the radial directions containing the caps.

\subsection{The halo} No  $T_{\rm e}$ determination is possible for this weak sub--system, whose electron density peaks at the inner edge 
($\simeq$250-300 cm$^{-3}$), and decreases outward. We assume $T_{\rm e}$(halo)$\simeq$10\,000 K. 

  \begin{figure*}
   \centering
\caption{Radial distribution of the physical conditions ($T_{\rm e}$ and $N_{\rm e}$) in selected directions of NGC 7009: top panel= the 
zvpc at PA=169$\degr$ (apparent minor axis), middle panel= the zvpc at PA=79$\degr$ (apparent major axis), bottom panel= the de-projected 
direction passing through the central star and the caps in PA=94$\degr$. Left ordinate scale:  $T_{\rm e}$[O III] (continuous line) and 
$T_{\rm e}$[N II] (squares). Right ordinate scale= $N_{\rm e}$[S II] (circles), $N_{\rm e}$[Ar IV] (dots) and $N_{\rm e}$(H$\alpha$) for three 
representative values of  $\epsilon_{\rm l}\times$r$_{\rm cspl}$$\times$D (dotted line= 5 arcsec kpc, short-dashed line= 10 arcsec kpc, and 
long-dashed line= 20 arcsec kpc).}
   \end{figure*}

\section{The nebular distance, size, mass and age}

The determination of the distance, D(NGC 7009), through the interstellar absorption--distance relation for the field stars (with accurate 
photometry, spectral type and luminosity class) furnishes poor results, due to the 
spread of the data and the modest nebular extinction (Lutz 1973, Gathier et al. 1986, Saurer 1995). 
A lower limit to the distance, D(NGC 7009)$\ge$1.0 kpc, is inferred from (D$\times \epsilon_{\rm l})_{\rm main\ shell}$ $\simeq$ 
(D$\times \epsilon_{\rm l})_{\rm outer\ shell}$ 
$\simeq$1.0 kpc (Sects. 4.2 and 4.5) and the assumption $\epsilon_{\rm l}$=1.

Our nebula is listed in more than two dozen catalogues of ``statistical'' distances (since O'Dell 1962), providing the 
following mean values: 
\begin{description}
\item[]$<$D$>$(Shklovsky)$\simeq$1.3($\pm$0.4) kpc
\item[]$<$D$>$(ionized mass--radius relation)$\simeq$1.0($\pm$0.5) kpc 
\item[]$<$D$>$(surface bright.--radius relation)$\simeq$1.0($\pm$0.5) kpc 
\item[]$<$D$>$(other methods)$\simeq$1.5($\pm$0.6) kpc. 
\end {description}

NGC 7009 being optically thin in most directions, $<$D$>$(Shklovsky) is expected to be more or less reliable; the same is valid for 
$<$D$>$(ionized mass--radius relation) and $<$D$>$(surface brightness--radius relation), but at a lower degree of confidence, due to 
the difficulty in defining the radius of such an elongated object. 

Note that $<$D$>$(Shklovsky) disagrees with both ``individual'' distances reported in the literature:
\begin{description}
\item[-] Liller (1965) and Liller et al. (1966) measured the angular expansion of the inner ring, the outer ring and the ansae on 
different epoch (1899, 1908 and 1961) plates taken with the 36-inch Crossley reflector, obtaining D(NGC 7009)$\simeq$0.6 kpc;
\item[-] Mendez et al. (1988) derived the mass, distance and luminosity of the central star by comparing the 
observed photospheric absorption-line profiles with the detailed profiles for a non--LTE model atmosphere. They give D(NGC 7009)$\simeq$2.5 kpc 
(later revised to 2.1 kpc by Mendez et al. 1992).
\end{description} 
Let us enter in some detail. 
According to Liller (1965) and Liller et al. (1966), the nebular material moves outward at a velocity proportional to the 
distance from the 
central star. They adopt the value $\frac{d\theta}{dt}$=0.7($\pm$0.3)$\times$10$^{-2}$ arcsec yr$^{-1}$ for the gas of the outer 
ring at 13 arcsec from the star; when combined with $V_{\rm exp}$=20.5 km s$^{-1}$ (Wilson 1950), this yields D(NGC 7009)=0.6($^{+0.4}_{-0.2}$) 
kpc, where:
\begin{equation}
  {\rm D(pc)}=\frac{0.211 V_{\rm exp} {\rm (km\ s^{-1})}}{\frac{d\theta}{dt} {\rm (arcsec\ yr^{-1})}}.
\end{equation}

After 1966 this distance has been widely adopted, and NGC 7009 is listed among the calibrators in a number of statistical studies on PNe. Now  
the kinematical results of Sect. 3 indicate that this must be revised. In fact, the expansion velocity of the gas in the outer shell 
at 13 arcsec from the star becomes $V_{\rm exp}\simeq$41 km s$^{-1}$ (from Eq. (4)), and the Liller (1965) and Liller et al. 
(1966) angular expansion, combined with Eq. (8), gives D(NGC 7009)=1.25($^{+0.9}_{-0.4}$) kpc. These authors measured also 
the angular expansion in the 
ansae, $\frac{d\theta}{dt}$=1.6($\pm$0.3)$\times$10$^{-2}$ arcsec yr$^{-1}$; from Eqs. (3) and (8) and r(ansae)$\simeq$25\arcsec\, we obtain 
$V_{\rm exp}$(ansae)$\simeq$100 km s$^{-1}$, and D(NGC 7009)=1.30($^{+0.7}_{-0.5}$) kpc (quite uncertain, since in this case we combine 
the poorly known expansion velocity of the gas with the angular expansion of the ionization front).  

The new values agree with $<$D$>$(Shklovsky), but are considerably lower than the ``gravity distance'' based on 
high-resolution spectra of the central star (Mendez et al. 1988, 1992). The literature contains contrasting reports on the general validity 
of the ``gravity distance'' for the central 
stars of PNe (see Pottasch 1996, Pottasch \& Acker 1998, Napiwotzki 1999, 2001). In the specific case of NGC 7009, Mendez et al. (1988, 
1992) obtained a quite large stellar mass (M$_*$=0.70 and 0.66 M$_\odot$, respectively; well above the canonical mean value 
for PNe nuclei, 0.60 M$_\odot$; Bl\"ocker 1995 and references therein), whose fast evolution appears incompatible with the observations 
(see below). Note that the same contrasting result occurs (in the opposite direction) when considering the distance 
suggested by Liller (1965) and Liller et al. (1966), i.e. 0.6 kpc: in this case the central star parameters correspond to an under--massive 
(M$_*\le$0.55 M$_\odot$), too slowly evolving post--AGB star. 

The definitive answer to the problem of the NGC 7009 distance will come from the angular expansion measured in first-- and second--epoch 
HST imaging, combined with the 
detailed spatio-kinematical results of Sect. 3. 
At present forty-four broad-band and interference filter WFPC2 images (often multiple exposures) of the nebula are contained in the HST 
archives, secured at five epochs covering the interval 1996.32 to 2001.36.  Two disturbing factors immediately appear: (i) the nebular size 
exceeds the field of the central (PC) camera in the WFPC2 mosaic; (ii) the telescope pointing changes from run to run, i.e. the nebula is 
located in different parts of the detector. For example, in the 1996.32 frames only the western region of NGC 7009 lies in the PC camera 
(pixel scale 0.046 arcsec pix$^{-1}$), whereas in the 2000.27 frames the whole nebula is in a lateral (WF) camera 
(pixel scale 0.100 arcsec pix$^{-1}$). 

All things considered (i.e. (1) the exciting star is the centring point of the first- and second-epoch observations, (2) the 
longer the temporal interval, the higher the angular expansion accuracy, and, mainly, (3) the frames  
``belong'' to someone else), we limit the  
analysis to the F555 images of the 1997.40 (program 
GO 6119; P.I. Howard Bond) and 2001.36 (program GO 8390; P.I. Arsen Hajian) runs. Multiple exposures are co-added, corrected for optical 
distortions, aligned and rotated using IRAF packages (Reed et al. 1999, Palen et al. 2002). 
We measure the apparent shift of the main shell close to the major axis (in the W-SW sector, at a distance of 11.2 arcsec from 
the central star), obtaining $\frac{d\theta}{dt}$=6.5($\pm2.0)\times$10$^{-3}$ arcsec yr$^{-1}$, and D(NGC 7009)=1450($^{+600}_{-400}$) pc 
(from Eqs. (3) and (8)).

This:
\begin{description}
\item[(a)] agrees with the results previously obtained from the Liller (1965) and Liller et al. (1966) ground-based angular expansion 
combined with the detailed spatio-kinematical model of Sect. 3;
\item[(b)] overcomes the (legitimate) doubts and criticisms on the expansion parallaxes of PNe (Sch\"onberner, private communication): this method 
compares matter velocities with pattern velocities, and can seriously underestimate the distance (up to a factor of two) when using an inaccurate 
spatio-kinematical model, as clearly shown by the Liller (1965) and Liller et al. (1966) angular expansion of NGC 7009, simply combined with the 
expansion velocity given by Wilson (1950).
\end{description}
In the following we will adopt D(NGC 7009)=1400 pc, corresponding to a distance from the galactic plane $|z|\simeq$780 pc. 
The resulting linear sizes of the different sub--systems are: r(main shell)$\simeq$0.038$\times$0.073 pc, r(outer shell)$\simeq$0.080 
pc, and r(ansae)$\simeq$0.161 pc. The local filling factor in the nebula is $\epsilon_{\rm l}\simeq$0.7 (Sects. 4.2 and 4.5), and the ionized mass 
M$_{\rm ion}$$\simeq$0.18($\pm$0.03) M$_\odot$ (obtained in various ways: from the H$\beta$ flux, the radio flux, and the observed $N_{\rm e}$ 
distribution; Aller 1984, Pottasch 1984, Osterbrock 1989, Turatto et al. 2002). M$_{\rm ion}$ is close to the 
total nebular mass, NGC 7009 being optically thin in most directions.

Concerning the kinematical age, t$_{\rm kin}\propto$ R/$V_{\rm exp}$, of the ``large-scale'' sub--systems, we have:
\begin{description}
 \item[-] t$_{\rm kin}$(main shell)$\simeq$1650 yr, 
\item[-] t$_{\rm kin}$(outer shell)$\simeq$2200 yr,
\item[-] t$_{\rm kin}$(halo)$\ge$8500 yr.
\end{description}

We can link these observational times with the theoretical ones coming from the most recent evolutionary models 
for PNe (Sch\"onberner et al. 1997, Corradi 
et al. 2000, Marigo et al. 2001, Villaver et al. 2002 and references therein); these predict the following chronological phases: AGB wind, superwind, 
transition time, photo-ionization, fast stellar wind, recombination and re--ionization (the last two phases mainly refer to a nebula 
powered by a massive, fast evolving star). Under the reasonable assumption that the observed halo of NGC 7009 corresponds to the AGB 
wind, we infer that t$_{\rm kin}$(halo) is an upper limit to the onset of the superwind, 
whereas t$_{\rm kin}$(outer shell) and t$_{\rm kin}$(main shell) are correlated to the ionization phase.

When taking into account the probable dynamical history of the gas, we obtain the rough chronological sequence of Table 4 
(the last column contains an estimate of the stellar temperature, T$_*$).
\begin{centering}
\begin{table}
\caption{NGC 7009: chronology desumed from the nebular kinematics}
\begin{tabular}{ccc}
\hline
\\
Time &Evolutionary phase& logT$_*$ \\
(yr)&& (K)
\\
\hline
\\
t$_0$-7000($\pm$1000)  & start of the superwind& 3.7\\
t$_0$-5000($\pm$1000)  & end of the superwind & 3.7 \\
t$_0$-2000($\pm$500)  & start of the photo-ionization& 4.4\\
t$_0$-1000($\pm$500)  & full ionization (main shell)& 4.7\\
t$_0$ &today& 4.9\\
\end{tabular}
\\
\\
\end{table}
\end{centering}
These evolutionary times, derived from the nebular kinematics, must be checked with those coming from the observed 
characteristics of the central star, as performed in the next section. 


\section{The central star}
The exciting star of NGC 7009 has m$_V$= 12.80($\pm$0.10) (Gathier \& Pottasch 1988, 
Tylenda et al. 1991, Ciardullo et al. 1999, Rubin et al. 2002) and  spectral type O (with H I, He I and He II absorption lines, Mendez et al. 1988).

The stellar magnitude, combined with both the total H$\beta$ nebular flux, log F(H$\beta$)$_{\rm obs}$=-9.78($\pm0.03$) 
mW$\times$m$^{-2}$ (Copetti 1990, Acker et al. 1992, Lame \& Pogge 1996, this paper), and the flux ratio F($\lambda$4686
$\AA$)/F(H$\beta$)=0.18($\pm0.03$) (Bohigas et al. 1994, Tylenda et al. 1994, Hyung \& Aller 1995a, b, Lame \& Pogge 1996, this paper), 
provides the following 
Zanstra temperatures: log(T$_{\rm Z}$H I)= 4.83($\pm0.05$) and log(T$_{\rm Z}$He II)=4.95($\pm0.05$). The Zanstra discrepancy confirms that 
the nebula is optically thin to the H Lyman continuum, i.e. T$_{\rm Z}$H I$<$T$_*$$\simeq$T$_{\rm Z}$He II, in agreement with the 
results by Gathier \& Pottasch (1988), Patriarchi et al. (1989) and Mendez et al. (1992). Moreover, from  D(NGC 7009)=1.4 kpc 
and c(H$\beta$)=0.15 we obtain log L$_*$/L$_\odot$(T$_{\rm Z}$H I)=3.50($\pm0.10$) and log L$_*$/L$_\odot$(T$_{\rm Z}$He II)=3.70($\pm0.10$) 
(using the bolometric corrections by Sch\"onberner 1981).

In summary, the central star of NGC 7009 is hot (logT$_*$$\simeq$4.95) and luminous (log L$_*$/L$_\odot$$\simeq$3.70). Its position in the 
log L$_*$--log T$_*$ diagram, compared with the theoretical evolutionary tracks 
by Sch\"onberner (1981, 1983), Iben (1984), 
Wood \& Faulkner (1986), Bl\"ocker \& Sch\"onberner (1990), Vassiliadis \& Wood (1994) and Bl\"ocker (1995), corresponds to a 0.60--0.61 
M$_\odot$ post--AGB star in the hydrogen--shell nuclear burning phase, whose evolutionary times fairly well agree with the kinematical results 
of Table 4. 

However, such a satisfactory picture, when analyzed in detail, presents a number of disturbing drifts:

I) the uncertainties in the model evolution for post-AGB are quite large (e.g. semi-empirical AGB mass--loss law, treatment of the residual 
hydrogen--rich envelope, energy release and neutrino losses in the fading to the  white dwarf region); 

II) the estimated mass--loss rate for the central star of NGC 7009    
spans the broad range 3.0$\times$10$^{-10}$ M$_\odot$ yr$^{-1}$ (Cerruti-Sola \& 
Perinotto 1985) to  1.0$\times$10$^{-8}$ M$_\odot$ yr$^{-1}$ (Bombeck et al. 1986); this is connected to the uncertainties 
in the stellar gravity, temperature and luminosity, and in the physical conditions, the ionic and chemical composition of the wind;

III) the diffuse X--ray emission of NGC 7009, arising in shocks at the interface between the fast stellar wind and the nebular shell, 
corresponds to a plasma at T$\simeq$1.8$\times$10$^6$ K (Guerrero et al. 2002), well below the post-shock temperature expected of the wind speed  
(V$_{\rm edge}$$\simeq$2700 km s$^{-1}$, Cerruti-Sola \& Perinotto 1985). Different temperature--moderating mechanisms have been introduced to overcome 
this discrepancy, such as: collimated outflows, adiabatic cooling, direct mixing and/or heat conduction between the shock-heated and the 
photo-ionized gas (for a review, see Soker \& Kastner 2003); 

IV) according to Guerrero et al. (2002), in NGC 7009 the thermal pressure of the X--ray emitting gas is large enough to push outward 
the internal, low-density nebular regions, whereas no evidence of acceleration appears in our echellograms, for the high excitation ``internal 
shell'' either. 
As suggested in Sect. 3, a  kinematical study of the (quite faint) [Ne V] emission at $\lambda$3425 $\rm\AA\/$ is highly desirable. 
The literature is desolately scanty in this field. We recall the classical paper of Wilson (1950), who measured $V_{\rm exp}$[Ne V] 
in eight bright PNe, obtaining $V_{\rm exp}$[Ne V]$\simeq$0 for five targets, and $V_{\rm exp}$[Ne V]=(the lowest value among the observed ions) 
for the remaining ones. Even though Meaburn \& Walsh (1980) reported the detection of  a 800 km s$^{-1}$ wide component to the [Ne V] 
$\lambda$3425 $\rm\AA\/$ 
emission of NGC 6302 (the first direct evidence of radiatively ionized fast stellar wind), this result is questioned by Casassus et al. (2000, 
2002): echellograms in the [Ne V] $\lambda$3425 $\rm\AA\/$ emission (R$\simeq$80\,000) and the IR coronal lines (R$\simeq$20\,000) of very 
high-excitation ions (up to 
[Mg VIII], IP 225 eV) show that all lines of NGC 6302 are very narrow, with no evidence of: (a) broad wings from fast stellar 
winds, (b) a wind-blown central cavity filled with a hot rarefied plasma (also see Oliva et al. 1996); 

V) the radial density profile of the hydrodynamical PNe models (Icke et al. 1992, Mellema 1997, Sch\"onberner et al. 1997, Dwarkadas 
\& Balick 1998, Garcia-Segura et al. 1999, Frank 1999, Corradi et al. 2000, Blackman et al. 2001, Soker \& Rappaport 2001, Marigo et al. 2001, 
Villaver et al. 2002) presents an internal, large, empty region (the hot bubble), whose bulldozer action on the nebular gas causes a very 
steep rise to the maximum density. This disagrees with the quite smooth internal radial density distribution observed in NGC 7009 (this paper), 
NGC 1501 (Ragazzoni et al. 2001), NGC 6565 
(Turatto et al. 2002), NGC 6818 (Benetti et al. 2003), and many other surveyed PNe. The odd thing is that, at present, the best match to 
the radial profiles of the observed PNe is provided by the hydrodynamical models neglecting (or limiting) the push effect by the fast wind 
on the innermost nebular regions; see, e. g., Bobrowsky \& Zipoy (1989) and Frank et al. (1990), though: (a) Marten \& Sch\"onberner (1991) 
argue that Bobrowsky \& Zipoy (1989) adopted unrealistic boundary conditions -no fossil AGB wind, no self-consistent central star evolution and inner 
rim pressure-, and Frank et al. (1990) did not consider the dynamical effects of photo-ionization; (b) according to  Mellema (1994), too low 
mass-loss rates (or velocities) in the fast post-AGB wind can cause the collapse of 
the nebula when ionization sets in (but see also Chevalier 1997).

The matter gets further complicated when introducing the ``small-scale'' sub--systems of NGC 7009 (the caps, the streams and the ansae). 
We will search for new input in the observed radial ionization structure and the photo-ionization model.

\section{The mean chemical and radial ionic abundances, and the photo-ionization model}

The mean ionic abundances of NGC 7009 are obtained from the line fluxes integrated over the spatial 
profile and the expansion velocity field, according to the critical analyses by Alexander \& Balick (1997) and Perinotto et al. (1998). 
Once the ionization  correcting
factors for the unobserved ionic stages have been considered (derived both empirically, Barker
1983, 1986, and from interpolation of theoretical nebular models, Shields
et al. 1981, Aller \& Czyzak 1983, Aller 1984, Osterbrock 1989), we find the chemical abundances listed on the bottom line of Table 5 
(to be compared with the estimates reported in the literature, also contained in the table). They are mean values representative of the main, 
bright nebula, and have been used (in Sect. 4) to derive F(H$\alpha)_{\rm zvpc}$ and $N_{\rm e}$(H$\alpha)$. 

\begin{centering}
\begin{table*}
\caption{NGC 7009: chemical abundances (relative to log H=12)}
\begin{tabular}{lccccccccc}
\hline
\\
Reference& Position& He& C& N& O &Ne&S&Cl&Ar\\
\\
\hline
\\
Liu et al. 1995&A&11.04$^{(*)}$&8.96$^{(*)}$&8.71$^{(*)}$&9.28$^{(*)}$&-&-&-&-\\
\\
Hyung \& Aller 1995a,b& B&10.99&8.43&8.26&8.73&8.13&7.14&5.26&6.40\\
& A&11.07&8.12&8.06&8.58&8.12&6.88&5.22&6.42\\
\\
&G&11.04&8.66&8.46&8.76&7.91&-&-&-\\
Kwitter \& Henry 1998&A&11.08&8.70&8.55&8.73&7.90&-&-&-\\
&E&11.08&8.61&8.48&8.76&8.09&-&-&-\\
&H&11.04&8.62&8.52&8.74&7.97&-&-&-\\
\\
Luo et al. 2001&A&-&-&-&-&8.24&-&-&-\\
&A&-&-&-&-&8.84$^{(*)}$&-&-&-\\ 
\\
&B&11.03-11.07&-&7.84-8.26&8.65-8.68&8.05&6.69-6.78&-&-\\
Gon\c calves et al. 2003&C+H&11.02-11.04&-&8.26-8.38&8.67-8.79&8.05-8.08&6.91-7.20&-&-\\
&S&11.03-11.09&-&7.79-7.93&8.48-8.81&7.98-8.05&6.64-6.80&-&-\\
&D+E&10.98-11.01&-&8.40-8.58&8.65-8.76&8.05-8.11&6.97-7.14&-&-\\
\\
This paper& F&11.05&8.2:&8.21&8.70&8.15&7.15&5.3:&6.40\\
\\
\\ 
\hline
\\
\end{tabular}
\begin{tabular}{llll}
(*) from optical recombination lines& A=main shell (minor axis)&D=West ansa&G=outer shell\\
&B=main shell (major axis)&E=East ansa&H=East cap\\
&C=West cap&F=overall nebula&S=streams\\
\\
\end{tabular}
\end{table*}
\end{centering}  

The detailed radial ionization structure of NGC 7009 is given by the zvpc of the different lines. Since the large H$\alpha$ broadening 
limits the accuracy of the 
F(H$\alpha)_{\rm zvpc}$ distribution, we adopt $\lambda$5007 $\rm\AA\/$ as 
reference emission, thus 
inferring the ionization structure relative to O$^{++}$ through:
\begin{equation}
\frac{X^{+a}}{O^{++}} = \frac{F(\lambda(X^{+b}))_{\rm zvpc}}{F(\lambda 5007\rm\AA)_{\rm zvpc}} f(T_{\rm e},N_{\rm e}),
\end{equation}
where a=b for the forbidden lines, and a=b+1 for the recombination lines (for details, see Benetti et al. 2003). 

\begin{figure*} \centering
\caption{The $\frac{X^{+i}}{O^{++}}$ radial ionization structure of NGC 7009 in the 
zvpc at PA=169$\degr$ (apparent minor axis; top panel), the zvpc at PA=79$\degr$ (apparent major axis; middle panel), and the de-projected 
direction passing through the central star and the caps in PA=94$\degr$ (bottom panel). Same orientation as Fig. 8.}  
\end{figure*}

The $\frac{X^{+i}}{O^{++}}$ profiles in the zvpc at PA=169$\degr$ (apparent minor axis), the zvpc at PA=79$\degr$ 
(apparent major axis), and the de--projected direction containing the central star and the caps in PA=94$\degr$ (see Sect. 4) are 
presented in Fig. 9. They confirm and/or enforce many of the observational results outlined in the previous sections. In particular:
\begin{description}
\item[-] along the minor axis (top panel) the nebula is optically thin to the 
UV stellar flux. Note, however, the presence of a dense, quasi--thick condensation in the northern sector of the outer shell (clearly visible in 
Fig. 1);
\item[-] along the major axis (middle panel) the internal shell, the main shell and the streams are high-excitation regions. The ionization 
quickly decreases in the ansae, whose outer edges are characterized by strong [O I] emission, i. e. they are neutral (or almost neutral) zones 
dominated by the resonant charge--exchange reaction O$^+$ + H$^0$$\getsto$O$^0$ + H$^+$ (Williams 1973);
\item[-] the same ionization drop occurs in the caps (bottom panel), which are shaded by the dense main shell. 
\end{description}
Note that in Fig. 9 both the ${\frac{He^+}{O^{++}}}$ and ${\frac{He^{++}}{O^{++}}}$ radial profiles tend to increase inward, suggesting an 
overabundance 
of He in the faint, innermost regions of NGC 7009. Although weakened by the low line fluxes and the He II $\lambda$4686 $\rm\AA\/$ fine-structure 
broadening, this agrees (at least qualitatively) with the different nebular morphology at $\lambda$4686 $\rm\AA\/$ and H$\alpha$ (Fig. 1). Let us 
enter in more detail, profiting of the superior spatial resolution of HST. 

Fig. 1 presents a large, well-defined stratification of the radiation: 
He$^{++}$ marks the highest-excitation layers, represented by 
the central hollow and the inner part of the main shell, whereas H$^+$ is emitted by the whole nebula, i.e. the central 
hollow, the entire main shell, the outer shell and the halo. In order to obtain the intrinsic $\frac{I(\lambda4686 \rm\AA\/)}{I(H\alpha)}$ ratio in 
the internal, high-excitation layers of NGC 7009, the H$\alpha$ contamination by the halo, the outer shell and the external part of 
the main shell must be removed in Fig. 1. 
This can be done in two ways:

(a) from  simple geometrical considerations, assuming spherical symmetry for the different emitting strata;

(b) using the F502 image as a marker of the He$^{++}$--poor layers, since IP(O$^{++}$)=54.9 eV$\simeq$IP(He$^+$)=54.4 eV; i.e. 
the [O III] $\lambda$5007 $\rm\AA\/$ line identifies the regions where He is single-ionized. 

Both these methods yield $\frac{I(\lambda4686 \rm\AA\/)}{I(H\alpha)}$$\ge$1.5  
in the innermost part of the He$^{++}$ shell and the central hollow of Fig. 1. 
Now, the $T_{\rm e}$ dependence of these two recombination lines is weak, and comparable. For a given 
$\frac{I(\lambda4686\rm\AA\/)}{I(H\alpha)}$ value, an increase of $T_{\rm e}$ implies a lower increase of $\frac{He^{++}}{H^+}$ 
(much lower for $T_{\rm e}$$\ge$15\,000 K), and an almost parallel emissivity decline (for example, when $T_{\rm e}$ doubles, 
going from 10\,000  to 20\,000 K, $\frac{He^{++}}{H^+}$ increases by 12$\%$, and the emissivities decrease 
by 53$\%$ for $\lambda$4686 $\rm\AA\/$, 
and 47$\%$ for H$\alpha$; Aller 1984, Pottasch 1984, Osterbrock 1989). 
Thus, in first approximation we can put $\frac{He^{++}}{H^+}\simeq0.23\times\frac{I(\lambda4686\rm\AA\/)}{I(H\alpha)}$. 
Moreover, in the internal, high-excitation regions $\frac{He^{++}}{H^+}\simeq\frac{He}{H}$.

We obtain ${\frac{He}{H}}$$\simeq$$\frac{He^{++}}{H^+}$$\ge$0.30 in the internal part of the He$^{++}$ shell and the central hollow of 
NGC 7009 (well above the mean helium abundance listed in Table 5).
Similar (but less precise) results come from our echellograms, when taking into account 
the $\lambda$4686 $\rm\AA\/$ and H$\alpha$ intensity profiles integrated over the whole expansion velocity field. 
All this, on the one hand supports the suspicion raised by the radial ionization structure, but on the other hand is far from conclusive, 
due to the large H$\alpha$ contamination by the halo, the outer shell and the external part of the main shell.
The literature is scarce and controversial: concerning NGC 7009, echellograms of the exciting star by Mendez et al. (1988) provide 
${\frac{He}{H}}$=0.05 at the stellar surface, and Guerrero et al. (2002) fit the X-ray spectrum of the central cavity 
using normal nebular abundances, whereas the extended X-ray spectrum of:

- NGC 6543 needs an overabundance of He by a factor of 60 with respect to 
the solar value (Chu et al. 2001; this result is questioned by Maness \& Vrtilek 2003);

- BD+30$\degr$3639 presents a significant continuum emission, suggestive of greatly enhanced He abundances (Kastner et al. 2000);

- NGC 7027 is overabundant in He, C, N, Mg and Si (Kastner et al. 2001). 

Moreover: 

- HST NICMOS imaging of NGC 5315 in different emission lines allowed Pottasch et al. (2002) to determine the helium abundance across 
the nebula: it is higher on the inside than on the outside, but not uniformly distributed (i.e. the He-rich gas appears in two opposite 
directions);

- both the kinematics and the radial ionization profile of NGC 40 suggest that the fast, H-depleted photospheric material ejected by the 
WC8 central star is gradually modifying the chemical composition of the innermost nebular regions (Sabbadin et al. 2000a).
\subsection{Photo-ionization model}
In this section the photo-ionization code CLOUDY (Ferland et al. 1998) is applied to a PN characterized by the same distance, 
gas distribution, mean chemical composition and exciting star parameters of NGC 7009. As usual (Sects. 4 and 7), we take into account 
different radial directions:
\begin{description}
\item[-] the zvpc in the S-SE sector at PA=169$\degr$ for the main shell, the outer shell and the halo along the apparent 
minor axis,
\item[-] the zvpc in the E-NE sector at PA=79$\degr$ for the internal shell, the main shell, the streams and the ansae along the (almost true) 
major axis,
\item[-] the de-projected direction containing the central star and the bright cap in the West sector at PA=94$\degr$.
\end{description}
\begin{figure*} \centering
\caption{Physical conditions and radial ionization structure in selected directions of the model nebula (CLOUDY), and the true nebula. 
Left column: zvpc in the S-SE sector at PA=169$\degr$ (apparent minor axis); central column: zvpc in the E-NE sector at PA=79$\degr$ 
(apparent major axis); 
right column: along the de--projected direction containing the central star and the West cap in PA=94$\degr$. The top and the middle 
panels concern the model nebula, and provide the physical conditions and the absolute fluxes in the main emissions, respectively; the bottom 
panels present the absolute radial flux distribution observed in the true nebula.}  
\end{figure*}

The input parameters of the model nebula are given in Table 6, whereas Fig. 10 presents the absolute radial flux distribution of the main 
emissions in the model nebula and the real nebula. We infer the following crucial points:
\begin{description} 
\item[(I)] all the spectral features observed in the sub-systems of NGC 7009 are fairly well explained by 
photo-ionization; an even better fit is obtained simply lowering $\epsilon_{\rm l}$(model) (0.68 instead of 0.70), and/or T$_*$(model) 
(85\,000 K instead of 90\,000 K). 

This: (a) agrees with the results by Balick et al. (1994, 1998), Hyung \& Aller (1995a, b)  and Gon\c calves et al. (2003), indicating that 
the nebula is mainly radiatively excited; 
(b) doesn't exclude that some shocked regions 
can be present in the ansae (as expected from the supersonic motion): following Dopita (1997), in the case of a shock illuminated by an external 
photo-ionizing source, the energetic flux of 
the ionizing photons overwhelms the mechanical energy flux through the shock;
\item[(II)] no N-enrichment need be postulated in the caps and the ansae to account for the local enhancement of the low 
excitation emissions. This feature is a mere consequence of the ionization drop: when increasing the gas distance from the 
central star, the ionization equilibrium moves towards lower states since the recombination rate is almost unchanged, whereas 
ionization is less frequent (more details are in Alexander \& Balick 1997 and Gruenwald \& Viegas 1998). 
At all radial directions of NGC 7009 (excluding the major axis) the main shell is a dense, optically thin layer degrading most of the UV 
stellar radiation, so that the mean latitude condensations in the outer shell (i.e. the caps) have to be optically thick. The main shell 
is weak along the major axis, and the UV stellar flux is here degraded in extended regions of moderate density (the streams), whose external, 
slightly denser edges (the ansae) yield the rapid plasma recombination.
\end{description}
\begin{table}
\caption{Input parameters for the model nebula (CLOUDY)}
\begin{tabular}{ll}
\hline
\\
Radial density profile &  Fig.~8 (cfr. Sect. 4)\\
\\
Chemical abundances:   & \\
~~ C, F, Cl, K, Ca & Hyung \& Aller (1995a,b)\\
~~ He, N, O, Ne, S, Ar & this paper\\
~~ other elements      & PN (CLOUDY default)\\
&\\
Dust                   & PN (CLOUDY default)\\
&\\
Local filling factor         & 0.7 \\
&\\
&blackbody distribution\\
Exciting star   & T$_*$=90\,000 K \\
& log L$_*$/L$_\odot$= 3.70\\
&\\
Distance& 1.4 kpc\\
\\
\hline
\end{tabular}
\end{table}

\section{The 3-D morpho-kinematical structure}
 
\begin{figure} \centering
\caption{
Stereoscopic structure of NGC~7009 for a rotation around the 
North-South  axis centered on the exciting star. Opaque reconstruction (at high and low flux cuts) for $\lambda$4686 $\rm\AA$
of He II, as seen from seven directions separated by 15$\degr$. The line of view
is given by ($\theta,\psi$), where $\theta$ is the zenith angle and $\psi$ the
azimuthal angle. Each horizontal pair represents a ``direct''
stereoscopic pair (i. e. parallel eyes), and the whole figure  provides six 3-D views of the nebula (at two flux cuts) in as
many directions, covering a right angle (for details, see Ragazzoni et al. 2001). The (0,0) images represent the rebuilt-nebula seen 
from the Earth (North is up and East to the left).
}  
\end{figure}

The 3--D reconstruction method for expanding nebulae was introduced by Sabbadin et al. (1985, 1987), and refined by Sabbadin et al. (2000a, b). 
In the case of the Saturn Nebula we select $\lambda$4686 $\rm\AA$ 
of He II, $\lambda$5007 $\rm\AA\/$ of [O III] and $\lambda$6584 $\rm\AA\/$ of [N II]
as markers of the high, mean and low-ionization regions,
respectively. 

The spectral images of the forbidden lines are de-convolved for seeing, spectral resolution and thermal motions, whereas also 
fine-structure is taken into account for $\lambda$4686 $\rm\AA$ of He II. The emissions of the main shell (+ the streams + the ansae) are de-projected 
through Eq. (3), and those of the outer shell (+ the caps) through Eq. (4), and later assembled by means of the 3-D rendering procedure described 
by Ragazzoni et al. (2001).

For reasons of space, in this paper we extract, present and discuss a limited number of frames, corresponding to a partial rotation around 
the North--South axis centered on the exciting star (i.e. almost perpendicular to the major axis). They provide a good picture of all aspects of 
the nebular complexity outlined in the previous sections. 
The complete series of nebular movies is available at  
{\bf http://web.pd.astro.it/sabbadin}; this WEB site dedicated to the spatial structure of the PNe must be considered as an integral 
(and important) part of the paper.
 
Figs. 11 to 13 contain the opaque reconstruction of NGC 7009 in He II, [O III] and [N II] (at different flux cuts) for a rotation of 90$\degr$ 
through the first Euler angle. The 
(0,0) images correspond to the Earth--nebula direction (North is up and East to the left).

The highest-excitation regions (Fig. 11) form an inhomogeneous ellipsoid, bright in two equatorial caps and open--ended along the major axis; 
at lower $\lambda$4686 $\rm\AA$ fluxes they merge into a closed structure with polar, diffuse layers. 

The opaque reconstruction in [O III] (Fig. 12) is given for three absolute flux cuts: log 
E($\lambda$5007 $\rm\AA$)= -16.82, -17.54 and -18.46 erg s$^{-1}$ cm$^{-3}$ (high, mean and low cuts, respectively). 
Since O/H=5.0$\times$10$^{-4}$$\simeq$O$^{++}$/H$^+$ (Table 5 in Sect. 7) and $N_{\rm e}\simeq$1.15$\times$N(H$^+$), they correspond to 
$N_{\rm e}$(high cut)$\simeq$5300 
cm$^{-3}$, $N_{\rm e}$(mean cut)$\simeq$2200 cm$^{-3}$, and $N_{\rm e}$(low cut)$\simeq$700 cm$^{-3}$ (assuming $T_{\rm e}$=10\,000 K and 
$\epsilon_{\rm l}$=1).  

The different sub--systems are clearly identifiable in these frames: the brightest regions of the main shell appear at high cut, the 
entire main shell, the caps and the densest parts of the outer shell at mean cut, and the overall nebular structure, including the 
streams and the ansae, at low cut. Moreover, the mean cut frames in Fig. 12 highlight the presence of a further (the eighth!) sub--system: 
an irregular, equatorial pseudo--ring within the outer shell. The newly discovered ring is barely visible (almost edge-on) in the H$\alpha$ 
and [N II] panels of Fig. 1; its most conspicuous feature is the quasi--thick knot in the northern sector at PA=169$\degr$ (Sects. 4.5 and 7).
 
In Fig. 14 the axes of the bright low-excitation sub--systems of NGC 7009, i.e. the ansae and the caps, are misaligned by 30$\degr$. 
At lower [N II] fluxes, the external parts of the main shell and the (just identified) equatorial pseudo-ring within the outer shell show up.
Please note the spatial correlation between the caps (Fig. 13, high cut) and the densest regions of the main shell (Fig. 12, high cut), 
confirming that the former do suffer strong shadowing by the latter (Sects. 3.4, 4, 7 and 7.1).

\begin{figure*} \centering
\caption{Opaque reconstruction of NGC 7009 at $\lambda$5007 $\rm\AA$ of [OIII] 
(high, mean and low flux cuts). Same scale as Fig. 11.
}  
\end{figure*}

The assembled, multi-color projection of NGC 7009 
for a rotation around the N--S direction (almost perpendicular to the major axis) is presented in Fig. 14, providing 
a representative sample of the nebular appearance when changing the line of view. 
The upper--right panel, (0,0),  corresponds to NGC 7009 as seen 
from the Earth (North is up and East to the left), to be compared with Fig. 1 and the HST multi-color images by Balick et al. 
(1998) and 
Hajian \&  Terzian at http://ad.usno.navy.mil/pne/gallery.html. 

Note the florilegium of silhouettes present in Fig. 14: actually, the Saturn Nebula covers most of the morphological classes, i.e. 
roundish to elliptical, to bipolar (according to Greig 1972, Stanghellini et al. 1993, Corradi \& Schwarz 1995, and  
Gorny et al. 1997). 
A similar variety of forms occurs for the other PNe so 
far analyzed with the 3-D procedure (i.e. 
NGC 6565 and NGC 6818; Turatto et al. 2002 and Benetti et al. 2003, respectively; see the movies at http://web.pd.astro.it/sabbadin). 
Such a dominant role of the perspective questions the validity of 
the statistical analyses based on a mere morphological classification of PNe, and stresses the need for a new code founded on 
``intrinsic, physical'' properties of the star+nebula system (instead of ``rough, apparent'' characteristics). 

\section{Discussion} 

\begin{figure*} \centering
\caption{Opaque reconstruction of NGC 7009 at $\lambda$6584 $\rm\AA$ of [NII] (high and low 
flux cuts). Same scale as Figs. 11 and 12.
}  
\end{figure*}

We have successfully applied the 3-D method (valid for all types of extended, expanding nebulae; Benetti et al. 2003 and references therein) 
to the complex PN NGC 7009, covered at high spatial and spectral resolutions with ESO NTT+EMMI. 
The Saturn Nebula lies at a distance of 1.4 kpc (age $\simeq$6000 yr, ionized mass $\simeq$0.18 M$_\odot$), and consists of four 
``large-scale'' sub--systems (the internal shell, the main shell, the outer shell and the halo) and as many ``small-scale'' ones (the streams, 
the caps, the ansae and the equatorial pseudo-ring), all interconnected and characterized by different morphology, excitation 
degree, physical conditions and kinematics. The four ``large-scale'' sub--systems and the streams are medium-to-high-excitation, 
optically thin regions, the caps and the ansae are thick, low-to-very 
low-excitation layers, whereas the equatorial pseudo-ring within the outer shell is quasi-thick (medium-to-low-excitation). The internal shell, 
the main shell, the streams and the ansae follow the expansion law 
$V_{\rm exp}$=4.0$\times$R\arcsec \,km s$^{-1}$, the outer shell, the caps and the pseudo-ring move at 
$V_{\rm exp}$= 3.15$\times$R\arcsec \,km s$^{-1}$, the 
halo expands at $V_{\rm exp}\simeq$10 km s$^{-1}$. 

The accurate definition of the 3-D ionization structure of NGC 7009 allows us to combine the radial distribution of the physical conditions 
and the line fluxes in the different sub-systems with the model 
profiles given by the photo-ionization code CLOUDY. We infer that all the observed features are explainable in terms 
of ionization by the central star, which is a hot (logT$_*$$\simeq$4.95) and luminous (log L$_*$/L$_\odot$$\simeq$3.70) 0.60--0.61 M$_\odot$ 
post--AGB star in the hydrogen-shell nuclear burning phase.

At this point we can compare the overall characteristics of NGC 7009 (star + nebula) with the two other PNe so far analyzed in detail with 
the 3-D methodology: 
\begin{description}
\item[NGC 6565](Turatto et al. 2002) is a young (2300 yr), optically thick ellipsoid embedded in a large 
cocoon of neutral, dusty gas. It is in a deep recombination phase (started about 400 yr ago), caused by the 
luminosity drop of the massive powering star (M$_*$$\ge$0.65M$_\odot$), which has almost reached the white 
dwarf region (log L$_*$/L$_\odot$$\simeq$2.0, log T$_*$$\simeq$5.08);
\item[NGC 6818](Benetti et al. 2003) is a young (3500 years), optically thin (quasi--thin in some directions) double shell  seen almost 
equator-on. It is at the start of the recombination phase following the luminosity decline of the 0.625 M$_\odot$ 
central star, which has recently exhausted the hydrogen shell nuclear burning and is rapidly moving toward the white dwarf domain 
(log T$_*\simeq$5.22 K; log L$_*$/L$_\odot\simeq$3.1).
\end{description}
NGC 7009 is found to be the oldest, but also the least evolved, member of the trio, confirming that the stellar mass is the main driver of PNe 
evolution (Bl\"ocker 1995 and references therein). 

\begin{figure*} \centering
\caption{Multi-color appearance of NGC 7009 (blue=He II, green=[O III], red=[N II]) for a rotation through the N--S axis centered on 
the exciting star. 
The upper-right panel, (0,0), corresponds to the re-built 
nebula as seen from the Earth (North is up and East to the left). Recall that 
projection($\theta,\psi$)=projection($\theta\pm180\degr,\psi\pm180\degr$).}  
\end{figure*}

When searching for more examples of NGC 7009--like PNe in the imagery catalogues (Acker et al. 1992, Schwarz et al. 1992, Manchado et al. 
1996, Gorny et al. 1999, and the HST archives), perspective effects play a leading role (as indicated by Fig. 14). We suspect this is also the 
case for NGC 2392, the famous 
``Eskimo Nebula''. According to Reay et al. (1983), O'Dell \& Ball (1985), O'Dell et al. (1990) and Phillips \& Cuesta (1999), it consists of 
an inner, bright, fast expanding ($\simeq$ 93 km s$^{-1}$) prolate spheroid (axial ratio 2:1) much denser in the equatorial regions, seen 
almost pole--on. 
A high-velocity stream ($\simeq$ 190 km s$^{-1}$) is formed at the open ends of the inner shell, probably connected to the accelerating 
effects by the fast stellar wind. NGC 2392 is embedded in a slow expanding ($\simeq$ 16 km s$^{-1}$) spheroid containing a fringe of fast 
($\simeq$ 75 km s$^{-1}$), low-excitation knots and filaments. Moreover, a collimated, jet--like bipolar flow with $V_{\rm exp}\simeq$200 km 
s$^{-1}$ has been detected by Gieseking et al. (1985) at PA=70$\degr$. The exciting star of the Eskimo Nebula has T$_{\rm Z}$H I=32\,700 K, 
T$_{\rm Z}$He II=73\,700 K, log L$_*$/L$_\odot$=3.99, and a mass--loss rate below 8.0$\times$10$^{-10}$ M$_\odot$ yr$^{-1}$, with 
V$_{\rm edge}$$\simeq$400--600 km s$^{-1}$ (assuming a distance of 1.5 kpc; Tinkler \& Lamers 2002). 

All this suggests the structural similarity of NGC 2392 and NGC 7009: the former is projected almost pole--on (see Fig. 9 in O'Dell \& Ball 
1985), the latter almost equator--on. 
Moreover, NGC 2392 is in a slightly earlier evolutionary phase, i.e. the main shell is 
fully ionized, but the outer shell is still thick (or quasi-thick) to the UV stellar flux, as supported by a preliminary analysis of the 
TNG+SARG (eight PA, R$\simeq$115\,000) and Asiago+Echelle (twelve PA, R$\simeq$25\,000) spectra of the Eskimo Nebula. 

Further examples of NGC 7009--like PNe are IC 4634, J 320, NGC 2371-2 and HB 4.

Now we are ready to combine the large amount of kinematical and physical information stored in the previous sections with the theoretical 
predictions (Bl\"ocker 1995, Mellema 1997, Capriotti 1998, Garcia-Segura et al. 1999, Marigo et al. 2001, Villaver et al. 2002 
and references therein), and 
sketch the evolution of an NGC 7009-like PN in a series of snapshots.

The first one, at time t$_0$-5000 yr, corresponds to the end of the superwind (this enhanced mass-loss phase lasts 1000--2000 yr). 
The central star parameters are: 
logT$_*$$\simeq$3.70 and logL$_*$/L$_\odot$ $\simeq$3.75. The newly born PN consists of about 0.15 M$_\odot$ of slowly expanding 
($V_{\rm exp}$=10--20 km s$^{-1}$) neutral gas (+ molecules + dust), whose structure is the result of two 
different superwind regimes: the older, almost spherical and isotropic, has been followed by a stronger, polar-depleted mass--loss. 
Moreover, a large cocoon of AGB wind embeds the nebula.

In the period t$_0$-5000 yr (logT$_*$$\simeq$3.70 and logL$_*$/L$_\odot$ $\simeq$3.75) to t$_0$-2000 yr (logT$_*$$\simeq$4.40 and 
logL$_*$/L$_\odot$$\simeq$3.75), corresponding to the transition phase, the UV stellar flux is still negligible, and the neutral 
proto--PN expands in a homologous 
fashion ($V_{\rm exp}$=10--20 km s$^{-1}$) maintaining the large pole-on to pole-off density contrast, also sustained by the 
moderate-velocity stellar wind, which prevents the gas backflow. According to Cant\'o et al. (1988) and Frank et al. (1996), the stellar wind 
can, eventually, be deflected poleward by interaction with the ellipsoidal nebula. 

Around t$_0$-2000 yr  (logT$_*$$\simeq$4.40 and logL$_*$/L$_\odot$$\simeq$3.75) photo-ionization starts, increasing the temperature 
(i.e. the kinetic energy) of the innermost nebular layers, which expand 
untill they reach pressure balance with the surroundings: inward expansion is prevented by the thermal pressure of the faster and faster 
stellar wind, whereas the external, adjacent, 
neutral regions are pushed outward, inducing fragmentation (i.e. Rayleigh--Taylor instabilities; Kahn \& Breitschwerdt 1990, 
Breitschwerdt \& Kahn 1990). 
Since the propagation velocity of the Str\"omgren radius depends on both the UV stellar flux and the density of 
the neutral gas, in all radial directions of NGC 7009 (except the major axis) the ionization front 
is trapped inside the dense main shell, whereas along the major axis it soon penetrates the outer shell (Mellema 1995). 

At t$_0$-1000 yr (logT$_*$$\simeq$4.70 and logL$_*$/L$_\odot$$\simeq$3.75) the ionization front includes the whole main shell, 
and in all directions (excluding the major axis) the gas pressure acts on the outer shell, which is still neutral. Along the 
major axis the ionization front has reached the external edge of the outer shell by now, further accelerating into the halo (shocks?). 
Note that in this phase 
the radiative cooling is inadequate, due to the increased velocity of the stellar wind, and an adiabatic shock forms 
at the interface with the nebular gas, squeezing the innermost layers of the main shell. Due to the compression of the main shell, the 
recombination rate increases, thus retarding the time at which the main shell is fully ionized (Breitschwerdt \& Kahn 1990).

In the last evolutionary snapshot, referring to t$_0$ (present time; logT$_*$$\simeq$4.95 and logL$_*$/L$_\odot$$\simeq$3.70), the whole nebula is 
optically thin to the UV stellar radiation, with the exception of:
\begin{description}
\item[-] the supersonic ionization fronts along the major axis, located well within the halo (i.e. the ansae),
\item[-] two strings of symmetric, mean latitude, thick condensations still present in the outer shell (the caps).
\end{description} 

What next?  At present the post-AGB star has reached the flat top 
of the UV flux (Bl\"ocker 1995 and references therein), and in the next millennium we do not expect great changes in the nebular structure, 
since the gas dilution due to 
expansion will balance the smooth decline of the ionizing photons. 
Later on, the stellar evolution 
presents a quick drop in luminosity (at the end of the hydrogen-shell nuclear burning), and the recombination processes 
will prevail on photo-ionization within the nebula. The recombination 
time scale being proportional to $\frac{1}{N{\rm e}}$ (Aller 1984, Pottasch 1984, Osterbrock 1989), the caps and the ansae will be the first to 
recombine (roughly at t$_0$+1100 yr; the ansae will showily contract), followed by the outer shell and the 
external layers of the main shell (at t$_0$+1200 yr), whereas the low-density halo will maintain a high-excitation degree for a long time. 
Beginning from t$_0$+2000 yr 
(logT$_*$$\simeq$5.1 and logL$_*$/L$_\odot$$\simeq$2.5), the slower and slower stellar decline, combined with the gas dilution due to 
expansion, will bring on the gradual re-ionization of the nebula (of course, starting with the outer layers of the main shell). 

Summing up: 
\begin{description}
\item[-] photo-ionization is the dominant process of the NGC 7009 shaping (but not the exclusive process, since the 
fast stellar wind models the innermost nebular regions; see Capriotti 1973, 1998). In particular, the 
peculiar excitation, structure and kinematics of the streams and the ansae recall the ``champagne-phase'' 
model (Tenorio-Tagle 1979) for H II regions with an outward decreasing density distribution: when the density gradient exceeds 
some critical value, the pressure gradients produce a weak R-type shock front that accelerates the gas to supersonic velocities 
(Franco et al. 1990, Garcia--Segura \& Franco 1996, Shu et al. 2002); according to Spitzer (1968), the speed of a weak R-type 
front is nearly an order of magnitude larger than the velocity of a I-type one (which is about the sound speed in the ionized 
region, i.e. $\simeq$10 km s$^{-1}$);
\item[-]  the outer shell and the caps of NGC 7009 are kinematically coeval (the latter simply representing local condensations within 
the former; Sects. 3.4 and 8); this is correlated to the last ionization phase of the nebula;
\item[-]  the three further co-moving sub--systems (i.e. the 
main shell, the streams and the ansae) can be considered kinematically coeval too, being associated to the start of the ionization process;
\item[-] the agent(s) and mechanism(s) producing the two-phase superwind ejection (first isotropic, then deficient along the major axis) are 
still nebulous: jets in the pre- and proto-PN phases, magnetic fields, binary central star, accretion disc around a close companion (for details, 
see Sahai \& Trauger 1998, Soker \& Rappaport 2000, Gardiner \& Frank 2001 and references therein);
\item[-] detailed 1-D and 2-D hydrodynamical simulation are needed (and highly desired) to improve/reject the rough picture here 
outlined.
\end{description}

Our simple evolutionary model (nearly a challenge), on the one hand interprets fairly well the observational characteristics of NGC 7009, 
on the other hand does not pretend to be the panacea for all microstructures of PNe. 
Gon\c calves et al. (2001)  list 50 objects with LIS (low ionization structures, including jets, 
jet-like systems, symmetrical and non--symmetrical knots), analyze the morphological and kinematical properties, and compare them with the 
current theoretical models.  An excellent review on FLIERs, jets and BRETs (bipolar rotating episodic jets) is provided by 
Balick \& Frank (2002, and references therein). 
Moreover, Corradi et al. (2000) and Benetti et al. (2003) demonstrate that the low ionization structures exhibited by NGC 2438 and NGC 6818 
can be explained in terms of a nebula in the recombination phase (caused by  the luminosity decline of the hot central star, 
which has exhausted the hydrogen shell nuclear burning and is moving toward the white dwarfs domain). 
We believe this is a not uncommon situation among PNe with LIS (cf. NGC 2440, NGC 2452, NGC 7293 and NGC 7354;  
Tylenda 1986). 

Let us focus on the ansae and the caps of NGC 7009. Strictly speaking, neither of them can be considered as genuine FLIERs, i.e. localized 
regions, whose characteristics (low excitation, N overabundance and peculiar motion with respect to the expansion law of the nebula) are 
suggestive of an exotic origin and/or evolution. In fact:
\begin{description}
\item[-] the caps are co-moving with their surroundings (i.e. the outer shell),
\item[-] the ansae follow the same kinematics of the main shell and the streams,
\item[-] the low excitation degree of both sub-systems can be explained in terms of ``normal'' nebular gas photo-ionized by the UV flux 
of the central star. 
\end{description}
Only in a broad sense are the ansae of NGC 7009 FLIERs, because of the peculiar motion with respect to the halo. 
In this case the streams become the prototype of a brand-new class of small-scale structures: the (sic!) 
LIFERs (large ionization, fast emitting regions).

Joking aside, the LIS in PNe remain a tangled skein. 
Given the variety of phenomenologies, and of physical and evolutional processes entering into the picture, we conclude that only a detailed 
analysis of a representative sample of targets can provide reliable and deep insights. To this end we have collected high-quality echellograms 
for a dozen objects with LIS.

Let us close the long journey dedicated to the Saturn Nebula listing the many problems left unsolved. 

Very likely the ansae expand supersonically, implying the presence of shocks (Sects. 3.1 and 7.1), but: (a) the mechanical 
energy flux through the 
shock is overwhelmed by the energetic flux of the photo-ionization (Dopita 1997); (b) the motion is almost tangential 
(i.e. in the plane of the sky), and the spatio-kinematical analysis based on the (small) radial velocity component inaccurate (Sects. 3.1 and 3.4). 
The same is valid 
for the streams. Deeper observations at even higher spatial and spectral resolutions (and in a wider spectral range) 
will shed light on this and on three more fields (the first two only grazed in the paper, the third not even touched): (i) the traverse, 
turbulent motions (i.e. the FWHM) in the streams and the ansae (Sects. 2 and 3.1), (ii) the spatio-kinematics of both the innermost, very 
high excitation layers (Sects. 3.1 and 6), and the extended, faint emissions surrounding the streams and the ansae (Sect. 3.3), 
and (iii) the discrepancy in the chemical composition of the gas obtained from optical recombination and collisionally excited lines 
(Hyung \& Aller 1995b, Liu et al. 1995, Luo et al. 2001). 

We infer (Sect. 2) that: (a) the baricentric radial velocity is the same in the outer shell, the caps, the ansae and the halo of NGC 7009, but these 
sub--systems are blue--shifted (by 3($\pm1$) km s$^{-1}$) with respect 
to the main shell; (b) the motion of the streams is hybrid; (c) the tip of both ansae is blue-ward. Can, all this, be connected to 
a binary nature of the central star (as suggested by the infrared photometry of Bentley 1989)? Does the different spatial orientation 
of the ansae and the caps (Sect. 8) support the precession hypothesis for the central star, eventually driven by a close companion 
(Bohigas et at. 1994)? 

Moreover: how can we reconcile the observed and the theoretical radial density profiles in PNe (Sects. 4 and 6)? Does the He 
overabundance in the innermost 
regions of NGC 7009 (Sect. 7) extend to other elements (and to other nebulae), implying a revision of the stellar mass-loss rates 
(so far obtained assuming nebular or 
solar abundances in the fast wind)? Do the detailed hydrodynamical simulations confirm/discard the naive evolutionary model here proposed (Sect. 9)
for the Saturn Nebula? What  
are the effects of a clumpy structure on the nebular evolution? And of magnetic fields?

To answer such a crescendo of questions, thorough 
investigations are required, both theoretical 
and observational, in a wide series of specific fields. Thus, mindful of Apelles's warning:``Cobbler, stick to thy last!'', we can only pass the baton 
to the specialists, and stop. 

\begin{acknowledgements}
This paper is dedicated to the memory of Lawrence Hugh Aller (1913--2003), the generous and tireless forty-niner, whose precious nuggets of 
learning have so much enriched our journey through the celestial nebulae. 

We would like to thank Denise Gon\c calves (the referee), Vincent Icke, Antonio Mampaso, Detlef Sch\"onberner and Noam Soker for their 
suggestions, criticisms and encouragements, and 
the support staff at the NTT (in particular, Olivier Hainaut) for assisting with the observations.
\end{acknowledgements}


\begin{thebibliography}{}


  \bibitem[1992]{c} Acker, A., Ochsenbein, F., Stenholm, B., et al. 1992, Strasbourg-ESO Catalogue of Galactic Planetary Nebulae, 
  ESO, Garching
  \bibitem[1997]{e} Alexander, J., \& Balick, B. 1997, AJ, 114, 713
  \bibitem[1984]{f} Aller, L. H. 1984, Physics of Gaseous Nebulae, Reidel, Dordrecht
  \bibitem[1983]{g} Aller, L. H., \& Czyzak, S. J. 1983, ApJS, 51, 211
  \bibitem[1938]{i} Baker, J. G., \& Menzel, D. H. 1938, ApJ, 88, 52
  \bibitem[1998]{klaswbz} Balick, B., Alexander, J., Hajian, A., et al. 1998, AJ, 116, 360
  \bibitem[2002]{kldsywe} Balick, B., \& Frank, A. 2002, ARA\&A, 40, 439
  \bibitem[1994]{xbnascghqty} Balick, B., Perinotto, M., Maccioni, A., et al. 1994, ApJ, 424, 800
  \bibitem[1993]{dkfu} Balick, B., Rugers, M., Terzian, Y., \& Chengalur, J. N. 1993, ApJ, 411, 778
  \bibitem[1986]{n} Barker, T. 1986, ApJ, 308, 314
  \bibitem[2003]{cn gddfhery} Benetti, S., Cappellaro, E., Ragazzoni, R., et al. 2003, A\&A, 400, 161
   \bibitem[1989]{nkzxcasjdfdff} Bentley, A. F. 1989, in IAU Symp. N. 131, Planetary Nebulae, ed. S. Torres-Peimbert, p. 312 
   \bibitem[1992]{cbvhjduew} Bianchi, L. 1992, A\&A, 260, 314
   \bibitem[1989]{kjsdjkm} Bianchi, L., Grewing, M., Barndtedt, J., \& Diesch, C. 1989, in IAU Symp. N. 131, Planetary Nebulae, ed. S. 
Torres-Peimbert, p. 182
  \bibitem[2001]{me} Blackman, E. G., Frank, A., \& Welch, C. 2001, ApJ, 546, 288
  \bibitem[1995]{o} Bl\"ocker, T. 1995, ApJ, 371, 217
  \bibitem[1990]{kw} Bl\"ocker, T., \& Sch\"onberner, D. 1990, A\&A, 240, L11 
  \bibitem[1989]{dfdfdf} Bobrowski, M., \& Zipoy, D. M. 1989, ApJ, 347, 307
   \bibitem[1994]{xcbn sdvjety} Bohigas, J., Lopez, J. A., \& Aguilar, L. 1994, A\&A, 291, 595
   \bibitem[1986]{bnkxchjsdtywe} Bombeck, G., K\"oppen, J., \& Bastian, U. 1986, in New Insights in Astrophysics (ESA SP-263), p. 287
   \bibitem[1990]{asjxhjdkfg} Breitschwerdt, D., \& Kahn, F. D. 1990, MNRAS, 244, 521
   \bibitem[1971]{r} Brocklehurst, M. 1971, MNRAS, 153, 471
   \bibitem[1988]{nmchjsjkf} Cant\'o, J., Tenorio-Tagle, G., \& Rozyczka, M. 1988, A\&A, 192, 287 
   \bibitem[1973]{cdddfdd} Capriotti, E. R. 1973, ApJ, 179, 495
   \bibitem[1998]{nbseeeeedf} Capriotti, E. R. 1998, RMxAC, 7, 173
   \bibitem[2000]{hjsdd} Casassus, S., Roche, P. F., \& Barlow, M. J. 2000, MNRAS, 314, 657
   \bibitem[2002]{jkdgfgjkasd} Casassus, S., Roche, P. F., Barlow, M. J., \& Binette, L. 2002, RMxAC, 12, 132
    \bibitem[1985]{ncbsgfsdr} Cerruti-Sola, M., \& Perinotto, M. 1985, ApJ, 291, 247
    \bibitem[1989]{ncbsgfldr} Cerruti-Sola, M., \& Perinotto, M. 1989, ApJ, 345, 339
 \bibitem[1997]{htrbgdfh} Chevalier, R. A. 1997, ApJ, 488, 263
   \bibitem[2001]{cdghweu} Chu, Y. -H., Guerrero, M. A., Gruendl, R. A. et al. 2001, ApJ, 553, L69
  \bibitem[1999]{xcdybyb} Ciardullo, R. B., Bond, H. E., Sipior, M. S., et al. 1999, AJ, 118, 480
   \bibitem[1990]{dfgfgftytyl} Copetti, M. V. F. 1990, PASP, 102, 77
  \bibitem[2000]{klsdidbd} Corradi, R. L. M., Gon\c calves, D. R., Villaver, E., et al. 2000, A\&A, 542, 861
   \bibitem[2000]{zxcczxccc} Corradi, R. L. M., Sch\"onberner, D., Steffen, M., \& Perinotto, M. 2000, A\&A, 345, 1071
  \bibitem[1995]{elsljkv} Corradi, R. L. M., \& Schwarz, H. E. 1995, A\&A, 293, 871
   \bibitem[1989]{nxcvgsdhof} Cristiani, S., Sabbadin, F., \& Ortolani, S. 1989, in IAU Symp. N. 131, Planetary Nebulae, ed. S. 
Torres-Peimbert, p. 191
   \bibitem[1997]{kjdsdfs} Dopita, M. 1997, ApJ, 485, 41
   \bibitem[1998]{nl} Dwarkadas, V. V., \& Balick, B. 1998, ApJ, 497, 267
  \bibitem[1998]{an} Ferland, G. J., Korista, K. T., Verner, D. A., et al. 1998, PASP, 110, 761
  \bibitem[1990]{zxcbnasddfgtr} Franco, J., Tenorio-Tagle, G., \& Bodenheimer, P. 1990, ApJ, 349, 126
  \bibitem[1999]{frg} Frank, A. 1999, NewAR, 43, 31
  \bibitem[1996]{klsnndrut} Frank, A., Balick, B., \& Livio, M. 1996, ApJ, 471, L53
  \bibitem[1990]{klsdrut} Frank, A., Balick, B., \& Riley, J. 1990, AJ, 100, 1903
  \bibitem[1996]{cbndffff} Garcia-Segura, G., \& Franco, J. 1996, ApJ, 469, 171
  \bibitem[1999]{tt} Garcia-Segura, G., Langer, N., Rozyczka, M., \& Franco, J. 1999, ApJ, 517, 767
  \bibitem[2001]{sdfaerdf} Gardiner, T. A., \& Frank, A. 2001, ApJ, 557, 250
  \bibitem[1988]{aq} Gathier, R., \& Pottasch, S. R. 1988, A\&A, 197, 266
  \bibitem[1986]{ap} Gathier, R., Pottasch, S. R., \& Pel, J. W. 1986, A\&A, 157, 171
  \bibitem[1985]{cdeddfg} Gieseking, F., Becker, I., \& Solf, J. 1985, ApJ, 295, L17
  \bibitem[2001]{hjuirkl} Gon\c calves, D. R., Corradi, R. L. M., \& Mampaso, A. 2001, ApJ, 547, 302 
  \bibitem[2003]{hjuirvv} Gon\c calves, D. R., Corradi, R. L. M., Mampaso, A., \& Perinotto, M. 2003, ApJ, 597, 975
  \bibitem[1999]{go} Gorny, S. K., Schwarz, H. E., Corradi, R. L. M., \& Van Winckel, H. 1999, A\&AS, 136, 145
  \bibitem[1997]{as} Gorny, S. K., Stasinska, G., \& Tylenda, R. 1997, A\&A, 318, 256
  \bibitem[1972]{at} Greig, W. E. 1972, A\&A, 18, 70
   \bibitem[1998]{fjkhgbv} Gruenwald, R., \& Viegas, S. M. 1998, ApJ, 501, 221
   \bibitem[2002]{xndywyw} Guerrero, M. A., Gruendl, R. A., \& Chu, Y. -H. 2002, A\&A, 387, L1
  \bibitem[1987]{av} Hummer, D. G., \& Storey, P. J. 1987, MNRAS, 224, 801
   \bibitem[1988]{sdfjkuiqwe} Hutsemekers, D., \& Surdej, J. 1988, A\&A, 219, 237
  \bibitem[1995]{hyy} Hyung, S., \& Aller, L. H. 1995a, MNRAS, 273, 958
  \bibitem[1995a]{hyy} Hyung, S., \& Aller, L. H. 1995b, MNRAS, 273, 973
  \bibitem[1984]{tz} Iben, I. Jr. 1984, ApJ, 277, 333
  \bibitem[1992]{lhjitrj} Icke, V., Balick, B., \& Frank, A. 1992, A\&A, 253, 224
   \bibitem[1990]{kdjkasiwer} Kahn, F. D., \& Breitschwerdt, D. 1990, MNRAS, 242, 505
   \bibitem[2000]{sdjkyuqwetyr} Kastner, J. H., Soker, N., Vrtilek, S. D., \& Dgani, R. 2000, ApJ, 545, L57
   \bibitem[2001]{kwetyqwecn} Kastner, J. H., Vrtilek, S. D., \& Soker, N. 2001, ApJ, 550, L189 
   \bibitem[1997]{fjktl} Keenan, F. P., McKenna, F. C., Bell, K. L., et al. 1997, ApJ, 487, 457
  \bibitem[1992]{bd} Kingsburgh, R. L., \& Barlow, M. J. 1992, MNRAS, 257, 317
   \bibitem[1998]{kldgit} Kwitter, K. B., \& Henry, R. B. C. 1998, ApJ, 493, 247
   \bibitem[1996]{nmxcvhjsdfsj} Lame, N. J., \& Pogge, R. W. 1996, AJ, 111, 2320
   \bibitem[1965]{m zcsdjweuu} Liller, W. 1965, PASP, 77, 25
   \bibitem[1966]{zxcbvadd} Liller, M. H., Welther, B. L., \& Liller, W. 1966, ApJ, 144, 280
    \bibitem[1993]{lu} Liu, X. -W., \& Danziger, I. J. 1993, MNRAS, 261, 465
   \bibitem[1995]{ludfdf} Liu, X. -W., Storey, P. J., Barlow, M. J., \& Clegg, R. E. S. 1995, MNRAS, 272, 369
   \bibitem[2001]{nmcdwe} Luo, S. -G., Liu, X. -W., \& Barlow, M. J. 2001, MNRAS, 326, 1049
   \bibitem[1973]{juilka} Lutz, J. H. 1973, ApJ, 181, 135
  \bibitem[1996]{od} Manchado, A., Guerrero, M. A., Stanghellini, L., \& Serra-Ricart, M. 1996, The IAC Morphological 
Catalog of Northern Planetary Nebulae, IAC
   \bibitem[2003]{dbnshwerui} Maness, H.L., \& Vrtilek, S. D. 2003, PASP, 115, 1002
   \bibitem[2001]{dfhkl;j} Marigo, P., Girardi, L., Groenewegen, M. A. T., \& Weiss, A. 2001, A\&A, 378, 958
   \bibitem[1991]{bksdwevscjl} Marten, H., \& Sch\"onberner, D. 1991, A\&A, 248, 590
   \bibitem[1998]{gl} Mathis, J. S., Torres-Peimbert, S., \& Peimbert, M. 1998, ApJ, 495, 328
  \bibitem[1980]{sdsdg} Meaburn, J., \& Walsh, J. R. 1980, MNRAS, 193, 631
  \bibitem[1988]{bo} Meatheringham, S. J., Wood, P. R., \& Faulkner, D. J. 1988, ApJ, 334, 862
  \bibitem[1994]{fjtbfhf} Mellema, G. 1994, A\&A, 290, 915
  \bibitem[1995]{uiogfdfs} Mellema, G. 1995, MNRAS, 277, 173
  \bibitem[1997]{zq} Mellema, G. 1997, A\&A, 321, L29
   \bibitem[1988]{xchj} Mendez, R. H., Kudritzki, R. P., Herrero, A., et al. 1988, A\&A, 190, 13
    \bibitem[1992]{xchjgkui} Mendez, R. H., Kudritzki, R. P., \& Herrero, A. 1992, A\&A, 260, 329   
   \bibitem[1988]{chjasdwe} Moreno-Corral, M., de La Fuente, E., \& Gutierrez, F. 1998, RMxAA, 34, 117
   \bibitem[1999]{mcccmgfjghj} Napiwotzki, R. 1999, AJ, 118, 488
   \bibitem[2001]{mvvcccmgfjghj} Napiwotzki, R. 2001, A\&A, 367, 973
   \bibitem[1962]{bq} O'Dell, C. R. 1962, ApJ, 135, 371
   \bibitem[1985]{cdghfdwet} O'Dell, C. R., \& Ball, M. E. 1985, ApJ, 289, 526
   \bibitem[1990]{nmcbdaqwe} O'Dell, C. R., Weiner, L. D., \& Chu, Y. -H. 1990, ApJ, 362, 226 
   \bibitem[1996]{cxnsdfjk} Oliva, E., Pasquali, A., \& Reconditi, M. 1996, A\&A, 305, L21 
   \bibitem[1989]{bs} Osterbrock, D. E. 1989, Astrophysics of Gaseous Nebulae and Active Galactic Nuclei, Mill Valley, 
CA Univ. Sci.
  \bibitem[2002]{klasfe} Palen, S., Balick, B., Hajian, A. R., et al. 2002, AJ, 123, 2666
   \bibitem[1989]{dgfgjkasf} Patriarchi, P., Cerruti-Sola, M., \& Perinotto, M. 1989, ApJ, 345, 327
   \bibitem[1981]{cvhjsdfe} Perinotto, M., \& Benvenuti, P. 1981, A\&A, 101, 88
   \bibitem[1998]{dbjkahjdjk} Perinotto, M., Kifonidis, K., Sch\"onberner, D., \& Marten, H. 1998, A\&A, 332, 1044
   \bibitem[1999]{zxcsd} Phillips, J. P., \& Cuesta, L. 1999, AJ, 118, 2929
  \bibitem[1984]{bz} Pottasch, S. R. 1984, Planetary Nebulae, a Study of Late Stages of Stellar Evolution, 
Reidel, Dordrecht
   \bibitem[1996]{zxcalkkl} Pottasch, S. R. 1996, A\&A, 307, 561
   \bibitem[1998]{zxcbjsdfwer} Pottasch, S. R., \& Acker, A. 1998, A\&A, 329, L5
   \bibitem[2002]{dftygh} Pottasch, S. R., Beintema, D. A., Bernard Salas, T., et al. 2002, A\&A, 393, 285 
  \bibitem[2001]{cc} Ragazzoni, R., Cappellaro, E., Benetti, S., et al. 2001, A\&A, 369, 1088 
   \bibitem[1985]{cnsvjdiero} Reay, N. K., \& Atherton, P. D. 1985, MNRAS, 215, 233 
   \bibitem[1983]{cnsvjdierof} Reay, N. K., Atherton, P. D., \& Taylor, K. 1983, MNRAS, 203, 1079 
  \bibitem[1999]{lwriubvcd} Reed, D. S., Balick, B., Hajian, A. R., et al. 1999, AJ, 118, 2430
   \bibitem[2003]{cadghdywer} Rubin, R. H., Bhatt, N. J., Dufour, R. J., et al. 2002, MNRAS, 334, 777
  \bibitem[2000]{cg} Sabbadin, F., Benetti, S., Cappellaro, E., \& Turatto, M. 2000b, A\&A, 361, 1112 
  \bibitem[1985]{jdsdfj} Sabbadin, F., Bianchini, A., Ortolani, S., \& Strafella, F. 1985, MNRAS, 217, 539
  \bibitem[2000]{ch} Sabbadin, F., Cappellaro, E., Benetti, S., et al. 2000a, A\&A, 355, 688 
  \bibitem[1987]{sdetq} Sabbadin, F., Cappellaro, E., \& Turatto, M. 1987, A\&A, 182, 305
  \bibitem[1998]{ss} Sahai, R., \& Trauger, J. T. 1998, AJ, 116, 1357
  \bibitem[1995]{sde} Saurer, W. 1995, A\&A, 297, 261
  \bibitem[1981]{cm} Sch\"onberner, D. 1981, A\&A, 103, 119
  \bibitem[1983]{cn} Sch\"onberner, D. 1983, ApJ, 272, 708
  \bibitem[1997]{ghdfsdghjs} Sch\"onberner, D., Steffen, M., \& Szczerba, R. 1997, Ap\&SS, 255, 459
  \bibitem[1992]{co} Schwarz, H. E., Corradi, R. L. M., \& Melnick, J. 1992, A\&AS, 96, 23
  \bibitem[1979]{sea} Seaton, M. J. 1979, MNRAS, 187, 73P
  \bibitem[1981]{cr} Shields, G. A., Aller, L. H., Keyes, C. D., \& Czyzak, S. J. 1981, ApJ, 248, 569
   \bibitem[2002]{dfdfhjhj} Shu, F. H., Lizano, S., Galli, D., et al. 2002, ApJ, 580, 969
   \bibitem[2003]{dhkfuier} Soker, N., \& Kastner, J. H. 2003, ApJ, 583, 368
   \bibitem[2000]{kr} Soker, N., \& Rappaport, S. 2000, ApJ, 538, 241
   \bibitem[2001]{kr} Soker, N., \& Rappaport, S. 2001, ApJ, 557, 256
  \bibitem[1978]{cs} Spitzer, L. Jr. 1978, Physical Processes in the Interstellar Medium, John Wiley \& 
Sons, New York
   \bibitem[1993]{ct} Stanghellini, L., Corradi, R. L. M., \& Schwarz, H. E. 1993, A\&A, 276, 463
   \bibitem[1979]{kdfkasdf} Tenorio-Tagle, G. 1979, A\&A, 71, 59
  \bibitem[2002]{zxcmsdk} Tinkler, C. M., \& Lamers, H. J. G. L. M. 2002, A\&A, 384, 987
  \bibitem[2002]{askdjwkeuf} Turatto, M., Cappellaro, E., Ragazzoni, R., et al. 2002, A\&A, 384, 1062 
  \bibitem[1986]{db} Tylenda, R. 1986, A\&A, 156, 217
   \bibitem[1991]{xczxcbdff} Tylenda, R., Acker, A., Raytchev, B., et al. 1991, A\&AS, 89, 77
  \bibitem[1992]{df} Tylenda, R., Acker, A., Stenholm, B., \& K\"oppen, J. 1992, A\&AS, 95, 337
   \bibitem[1994]{mvjjgggy} Tylenda, R., Stasinska, G., Acker, A., \& Stenholm, B. 1994, A\&AS, 106, 559
  \bibitem[1994]{rh} Vassiliadis, E., \& Wood, P. R. 1994, ApJS, 92, 125
  \bibitem[2002]{dvhbfhn} Villaver, E., Manchado, A., \& Garcia-Segura, G. 2002, ApJ, 581, 1204
  \bibitem[1968]{ff} Weedman, D. W. 1968, ApJ, 153, 49
  \bibitem[1973]{dl} Williams, R. E. 1973, MNRAS, 164, 111
  \bibitem[1950]{dm} Wilson, O. C. 1950, ApJ, 111, 279
  \bibitem[1986]{zh} Wood, P. R., \& Faulkner, D. J. 1986, ApJ, 307, 659

\end{thebibliography}
\end{document}